\documentclass[nohyper,12pt,letterpaper]{JHEP3}
\usepackage{epsfig,amsfonts, textcomp, amssymb}
\usepackage{amsmath}

\newcommand{\ben}{\begin{eqnarray}\displaystyle}
\newcommand{\een}{\end{eqnarray}}
\newcommand{\be}{\begin{equation}}
\newcommand{\ee}{\end{equation}}
\newcommand{\lb}{\left (}
\newcommand{\rb}{\right )}
\newcommand{\ltb}{\left [}
\newcommand{\rtb}{\right ]}
\newcommand{\ra}{\rightarrow}

\newcommand{\nn}{\nonumber}
\newcommand{\nt}{{1\over 16 \pi G_5}}

\newcommand{\ep}{\epsilon}

\newcommand{\cA}{{\cal A}}

\newcommand{\ph}{\phi(r,k)}
\newcommand{\php}{\phi'(r,k)}
\newcommand{\cph}{\phi(r,-k)}
\newcommand{\cphp}{\phi'(r,-k)}

\title{\center{Moduli and electromagnetic black brane}\\
\center{holography}}

\author{\hspace{1.5cm} Dumitru Astefanesei,$^a$ Nabamita Banerjee,$^b$ and Suvankar Dutta$^c$\\

$^a$Max-Planck-Institut f\"ur Gravitationsphysik,
Albert-Einstein-Institut, 14476 Golm, Germany\\
$^a$Erwin Scr\"odinger Inst., 1090 Wien, Austria \\
$^b$ Institute of Theoretical Physics, Utrecht University, Leuvenlaan 4
3584 CE Utrecht, The Netherlands\\
$^c$ Department of Physics,
Swansea University,
Singleton Park,
Swansea, SA2 8PP, UK\\

\\E-mail: \email{dumitru@aei.mpg.de, N.Banerjee@uu.nl,pysd@swan.ac.uk}}

\abstract{ We investigate the thermodynamic and hydrodynamic properties of $4$-dimensional
gauge theories with finite electric charge density in the presence of a
constant magnetic field. Their gravity duals are planar magnetically and
electrically charged AdS black holes in theories
that contain a gauge Chern-Simons term. We present a careful analysis of
the near horizon geometry of these black branes at finite and zero
temperature for the case of a scalar field non-minimally coupled to
the electromagnetic field. With the knowledge of the near horizon data, we obtain
analytic expressions for the shear viscosity coefficient and entropy
density, and also study the effect of a generic set of four derivative
interactions on their ratio. We also comment on the attractor flows of the
extremal solutions.}

\keywords{AdS/CFT, transport coefficients, higher derivative gravity,
attractor mechanism}

\preprint{AEI-2010-135}

\begin{document}{\vskip 1cm}

\newpage
\section{Introduction}
\label{intro}

The exploration of the phase diagram of Quantum Chromodynamics (QCD)
(see, e.g., \cite{Rajagopal:2000wf}) is a very active area of ongoing
research in both, experimental and theoretical physics.

It is well known that the zero temperature ground state of QCD (at
normal nuclear densities) has two important features: chiral symmetry
is broken spontaneously and the color charges are confined to the
interior of the individual hadrons (the scale is about 1$fermi$).
However, the studies in lattice QCD predicted that, beyond
a temperature of about $150MeV$ (in fact, the critical temperature is
$T_c\sim 200MeV$), the hadronic matter undergoes a
transition from the confined phase to a deconfined color phase
(of quarks and gluons).\footnote{Also, by increasing the
temperature and/or density the QCD vacuum restores its chiral
invariance.}

Since the transition temperature is not extremely high, it was expected
that this phase of matter can even be produced in the laboratory. Indeed,
the experimental data obtained at the Relativistic Heavy Ion Collider
(RHIC) have confirmed the `theoretical' expectation.

What came as a big surprise, though, is that the experimental data
indicate that the {\it quark-gluon plasma} is a  new state of thermalized
matter, which exhibits almost {\it ideal hydrodynamic} behaviour. The
early thermalization and a very small viscosity to entropy density
area ($\eta/s$) are characteristics of a strongly interacting system --- at
weak coupling ($g\ll 1$), the equilibration time and ratio $\eta/s$ are
proportional with the mean free path ($\sim 1/g^4$) and so they are
parametrically large.

In other words, the experimental data are consistent with the interpretation
of the new state as a strongly interacting plasma (see, e.g.,
\cite{Kolb:2003dz} and references therein). This new phase of nuclear matter
(the state of deconfinement super-hot hadronic matter) is known as the
strongly coupled quark-gluon plasma (sQGP).

Obviously, the perturbation theory is not useful to investigate
the properties of sQGP. Also, the lattice QCD study is not suitable
for computing the dynamical quantities of sQGP. Interestingly enough,
valuable guidance for understanding the near perfect fluidity observed
at RHIC was obtained from the Anti-deSitter(AdS)/conformal field
theory(CFT) duality \cite{Maldacena:1997re}.\footnote{A concise review of
{\it phenomenological} problems that can not be solved by the standard
field theoretic approach, but for which the AdS/CFT duality may be
useful, can be found in \cite{Zakharov:2007zzb}.}

It is important to emphasize that the gravity dual of QCD
is not known. However, the QCD matter of interest is in a
deconfined phase for which the `conformal anomaly' (due to the
running of the coupling) appears to be relatively small. Therefore,
since QCD is approximately conformal at sufficiently
large energies, one expects some similarities with some of the gauge
theories that can be studied by using the AdS/CFT duality.

Indeed, a study of quantum field theories (QFT) with a gravity dual
revealed the fact that, when heated up to finite temperature, these
QFTs behave hydrodynamically (see, e.g., \cite{Son:2009zz}
and references therein and, also, \cite{Bhattacharyya:2008jc})
at large distances and time scales.

The holographic hydrodynamics (the first attempt
to study hydrodynamics via AdS/CFT was \cite{pss}) is
an important tool for understanding some properties of strongly
coupled quantum field theories in terms of AdS black holes physics. A notorious
example is the computation
of the ratio of the shear viscosity ($\eta$) to the entropy density
($s$). It was proposed in \cite{kovtun} that there is an universal
viscosity bound\footnote{However, there are known examples for which
the bound is violated (see \cite{Sinha:2009ev} and references therein).},
which is also satisfied by sQGP:
\be
\frac{\eta}{s} \geq  \frac{\hslash}{4\pi k_B}.
\ee

Kubo formula relates the shear viscosity to the two-point function
of energy momentum tensor in zero frequency limit (see Appendix A for
a review). From the field/operator correspondence of
AdS/CFT duality, we know that the energy momentum tensor
of boundary field theory is sourced by bulk graviton excitations.
In \cite{pss}, the authors have considered graviton excitations
polarized parallel to the black brane, which are moving transverse
to it. They found that the graviton absorption cross section is related
to the shear viscosity coefficient.

In this paper, we use holographic techniques to investigate the
properties of $4$-dimensional gauge theories with finite electric
charge density in the presence of a constant magnetic
field. Strong magnetic fields are created in heavy ion
collisions at RHIC, and some models were studied in \cite{Fukushima:2010fe}.

A study of black branes
solutions of Einstein-Maxwell AdS gravity with a gauge Chern-Simons
term can be found in \cite{D'Hoker:2009bc, D'Hoker:2010rz}. We use the
entropy function to carefully study the near horizon geometry of
these black branes at zero temperature and discuss different branches
with finite area horizons. This allows us to obtain an analytic expression
for the entropy density that supports the numerical analysis of
\cite{D'Hoker:2010rz} (though, we find a larger class of near
horizon geometries). That is, there is a critical value of the magnetic
field for which the entropy vanishes.

We also extend the work of \cite{D'Hoker:2009bc} by including a
modulus, which is exponentially coupled to the gauge field
kinetic term in the action. This kind of coupling appears in
consistent truncations of type $IIB$ supergravity on Sasaki-Einstein
manifolds \cite{Liu:2010sa}. In the extremal limit, we argue
that, due to the attractor mechanism, the near horizon geometry
is stable under scalar fields fluctuations.

More importantly, unlike \cite{D'Hoker:2009bc}, our main focus is
on the hydrodynamics properties of this system. We use the membrane
paradigm approach \cite{liu} and its higher derivative generalization
\cite{ns1} (see, also, \cite{cai1}) to compute the shear viscosity (related
work for finite chemical potential can be found in \cite{mps, cai2}). In
particular, we study the effect of the moduli (in the presence of a
generic set of four-derivative interactions) on the viscosity bound.

Our results are given in terms of near horizon data, and not in
terms of the field theory parameters that would be measured at the $AdS_5$
boundary. For a computation of the viscosity of our system, that is enough.
However, there is a non-trivial flow for other transport coefficients,
e.g. the conductivity. Also, it will be interesting to express the entropy
in terms of asymptotic parameters --- in \cite{D'Hoker:2010rz} this problem
was solved analytically (for vanishing scalar field). We will present some
of these results in the forthcoming work \cite{DNS2, DNS1}.

An overview of the paper is as follows: In Section \ref{setup},
we present the bulk action, set-up the conventions for our model, and
provide some useful holographic data.
In Section 3, we carefully study the near horizon geometry of
electromagnetic black branes at zero and finite temperature. In the
extremal limit, we use the entropy function and present exact
near horizon geometry solutions. In Section \ref{etabys}, we
compute the shear viscosity to entropy density in the presence of a generic
set of four derivative interactions. Finally, we conclude with
a discussion of our results and present some future directions.
The appendices contain supporting material.


\section{General set-up}\label{setup}

The $AdS/CFT$ duality provides a concrete relation between the
regular black brane solutions in $AdS_5$ and the hydrodynamic regime
of strongly coupled $4$-dimensional quantum field theories.

In this section, we fix the conventions for our model and present
a brief review of holography relevant to our work. We follow closely
\cite{D'Hoker:2009bc}, though our model is different because we also
consider scalar fields. In this way, it is straightforwardly to
compare some of the results. However, one important difference is
that we do not rescale the horizon radius to one, as in
\cite{D'Hoker:2009bc} --- the physics in the presence of a scalar
field is unambiguous if we do not use this rescaling.

\subsection{The model} \label{model}
We will focus on a 5-dimensional theory of gravity coupled
to a massless scalar and an abelian electromagnetic field
whose action is\footnote{We also have to add counterterms
to regularize the action --- for (\ref{action1}), the
counterterms are given in \cite{Taylor:2000xw} (see, also,
the nice review \cite{Skenderis:2002wp} and references therein
for a more detailed discussion).}
\be
\label{action1}
S_{{\rm EM}} = \frac{1}{16 \pi G_5} \int d^5x \sqrt{-g}
\left( R + {12} - e^{\alpha \varphi(r)} F_{\mu\nu} F^{\mu\nu} -\partial_{\mu}\varphi \partial^{\mu}\varphi \right) + S_{{\rm CS}},
\ee
\be
\label{yb}
S_{{\rm CS}} = {{\zeta\over 3} \over 16\pi G_5} \int  d^5x\ \epsilon^{\mu\nu\rho\sigma\gamma}A_{\mu} F_{\nu\rho} F_{\sigma\gamma}.
\ee
In this paper, we consider a constant moduli potential,
$V(\phi)=2\Lambda=-12/l^2$, and also fix the radius of $AdS$ to be $l=1$.

Since the equations of motion for the gauge field simplify, we choose
the constant in front of the Chern-Simons term to be $\zeta/3$. The action
(\ref{action1}) with various values for $\alpha$ resembles (truncated)
actions obtained in string compactifications \cite{Liu:2010sa}. The coupling $\zeta$
captures the strength of the anomaly of the boundary current.

The equations of motion for the metric, scalar, and electromagnetic
field ($F_{\mu\nu}=\partial_{\mu}A_{\nu}-\partial_{\nu}A_{\mu}$) are
\be
\label{einstein}
R_{\mu\nu}+4 g_{\mu\nu}+e^{\alpha \varphi(r)}\lb {1\over 3}
F^2 g_{\mu\nu}-2 F_{\mu\rho}F_{\nu\sigma}g^{\rho\sigma}\rb-\partial_{\mu}\varphi \partial_{\nu}\varphi=0\ ,
\ee
\be
\label{scalar}
\frac{1}{\sqrt{-g}}\partial_{\mu}(\sqrt{-g}\partial^{\mu}\varphi)
  = -\frac{1}{2}\alpha e^{\alpha \varphi(r)}F_{\mu\nu} F^{\mu\nu}\ ,
\ee
\be
\label{gauge}
e^{\alpha\varphi(r)}\partial_{\nu}\lb \sqrt{-g}F^{\nu\mu}\rb + {\zeta\over 4} \epsilon^{\mu\rho\sigma\gamma\delta}F_{\rho\sigma}F_{\gamma\delta}=0\ ,
\ee
where we have varied the scalar and the electromagnetic field
independently. The Bianchi identities for the gauge field are
$F_{[\mu\nu;\lambda]}=0$.

Since we are interested in a theory for which the Chern-Simons term
has a non-trivial contribution, we consider the following ansatz for the
gauge field:\footnote{Please note that our $P'(r)$ is the same as $P(r)$ in
\cite{D'Hoker:2009bc} and our $Z(r)$ in the ansatz of the metric
(\ref{anz2}) is their $C(r)$.}
\be
A=E(r) dt - {B\over 2} y dx + {B\over 2} x dy - P(r) dz\ .
\ee
The magnetic field, $B$, is fixed to be constant by the Bianchi
identities. Thus, the field strength is \footnote{Our convention
for the coordinates is $(r,t,x,y,z)$.}
\be
\label{Fansatz}
F=\left(
\begin{array}{ccccc}
 0 & E'(r) & 0 & 0 & -P'(r) \\
 -E'(r) & 0 & 0 & 0 & 0 \\
 0 & 0 & 0 & B & 0 \\
 0 & 0 & -B & 0 & 0 \\
 P'(r) & 0 & 0 & 0 & 0
\end{array}
\right)\ ,
\ee
where $'$ denotes derivatives with respect to $r$.

Our analysis is on time-independent black hole solutions and so we consider
the following ansatz for the metric
\be
\label{anz2}
ds^2 = {dr^2 \over U(r)} - U(r) dt^2 + e^{2V(r)} \left ( dx^2 + dy^2 \right  )
+ e^{2 W(r)} \left ( dz + Z (r) dt \right )^2\ ,
\ee
which is compatible with the symmetries of the problem.

In this case, the horizon is located at (the biggest
root of) $U(r_h)=0$ and the temperature can be easily computed on the
Euclidean section --- we obtain
\be\label{temp}
T= {U'(r_h)\over 4\pi}\ .
\ee

By using the metric ansatz (\ref{anz2}), we can rewrite the Maxwell
equations as
\be
\label{ME1}
[Q(r) e^{2 V(r)+W(r)+\alpha  \varphi (r)}]' - 2  \zeta B P'(r)=0,
\ee
\be
\label{ME2}
[e^{2 V(r)+W(r)+\alpha  \varphi (r)} \left(U(r) e^{-2 W(r)}
   P'(r)-Q(r) Z(r)\right)]' - 2 \zeta B E'(r)=0
\ee
where
\be
Q(r) = E'(r) + Z(r) P'(r).
\ee

It is easier to work with combinations of Einstein equations rather
than using directly (\ref{einstein}). First, we extract the expressions
of second derivatives of the functions that characterize the metric in
the following way: we obtain $W''(r)$ from $(rr)$- , $V''(r)$ from
$(xx)$- , and $U''(r)$ from $(zz)$-component of Einstein equations.

Let us now consider the $(tz)$-component of Einstein equations in which
we replace $W''(r), V''(r),$ and $U''(r)$ --- we obtain
\be
\label{E4}
e^{2 W(r)} \left[2 V'(r) Z'(r)+3 W'(r) Z'(r)+Z''(r)\right]-4
   Q(r) P'(r) e^{\alpha  \varphi (r)}=0.
\ee
%

An important observation, which can be drawn by studying the system of equations
(\ref{ME1})--(\ref{E4}), is that a non-zero magnetic field is not compatible
with a constant function $Z(r)$ (and, also, $P(r)$).

The other (independent) combinations of Einstein equations
are obtained as follows:
by replacing $U''(r)$ in the $(rr)$-component of Einstein equations we get
\ben\label{E1}
&& 2 B^2 e^{\alpha  \varphi (r)-4 V(r)}+2 Q(r)^2 e^{\alpha  \varphi
   (r)}+U(r) [2 V'(r)+W'(r)]^2\nn \\
&&+[U(r) \left(2
   V'(r)+W'(r)\right)]'+\frac{1}{2} e^{2 W(r)} Z'(r)^2-12=0
\een
by replacing $W''(r)$ in the $(zz)$-component of Einstein equations
\ben\label{E3}
&&-4 Q(r)^2 e^{\alpha  \varphi (r)}+U''(r)+U'(r) [2
   V'(r)-W'(r)]-2 e^{2 W(r)} Z'(r)^2\nn \\
&&+2 U(r) [2 V''(r)-2 V'(r) W'(r)+2
   V'(r)^2+\varphi '(r)^2]=0
\een
and the last one is, in fact, the $(xx)$-component of Einstein equations
\ben\label{E2}
&&e^{2 W(r)} [-4 B^2 e^{\alpha  \varphi (r)}-3 e^{4 V(r)}
   U'(r) V'(r)+12 e^{4 V(r)}]-2 Q(r)^2
   e^{4 V(r)+2 W(r)+\alpha  \varphi (r)}\nn \\
&&+U(r) e^{4 V(r)} [2
   P'(r)^2 e^{\alpha  \varphi (r)}-3 e^{2 W(r)}
   \left(V''(r)+V'(r) W'(r)+2 V'(r)^2\right)]=0.
\een
We also use the ansatz of the metric in the equation of motion for
the scalar (\ref{scalar}) and so this equation becomes
\be
\label{SC}
[U(r) e^{2 V(r)+W(r)} \varphi '(r)]'+ \alpha  e^{2 V(r)+W(r)+\alpha  \varphi (r)} [B^2 e^{-4
   V(r)}+U(r) e^{-2 W(r)} P'(r)^2-Q(r)^2]=0.
\ee

Due to the non-trivial coupling between the scalar and gauge fields,
the equation (\ref{scalar}) has a non-trivial right hand side. The
non-vanishing electromagnetic field may also be understood as a source
for the scalar field. Thus, the scalar charge is determined by the
electric and magnetic charges and so it is not an independent parameter
that characterizes the system --- this charge plays an important
role when the asymptotic value of the scalar is not fixed (see
\cite{Gibbons:1996af}).

We would like to conclude this section with an observation on the
Hamiltonian constraint. A vanishing Hamiltonian is a characteristic
feature of any theory that is invariant under arbitrary coordinate
transformations --- for our system, we can obtain a first order
differential equation by replacing $W''(r), V''(r),$ and $U''(r)$
in the $(tt)$-component of Einstein equations.

The Hamiltonian constraint, which can be enforced as an initial
condition, has the following expression:
\ben\label{CON}
&& 2 B^2 e^{\alpha  \varphi (r)-4 V(r)}+U(r)[-2 P'(r)^2
   e^{\alpha  \varphi (r)-2 W(r)}+4 V'(r) W'(r)+2
   V'(r)^2- \\ \nn
&& -\varphi '(r)^2]+2 Q(r)^2 e^{\alpha  \varphi
   (r)}+2 U'(r) V'(r)+U'(r) W'(r)+\frac{1}{2} e^{2 W(r)}
   Z'(r)^2-12=0 \ .
\een

\subsection{Holographic data}
\label{holographicdata}

The AdS/CFT correspondence \cite{Maldacena:1997re} is a concrete example of open/closed
string duality. Remarkably, at strong coupling $\mathcal{N}=4$ SYM
is best described by classical supergravity --- in this section,
we closely follow the reviews \cite{Skenderis:2002wp,Nastase:2007kj}.

The observables in the field theory side of the duality are the
correlation functions of gauge invariant operators, which are
composites of the elementary fields. Any supergravity field, $\Phi$,
corresponds to an operator, $\mathcal{O}$, in the (boundary) field
theory --- in particular, bulk gauge fields correspond to boundary
symmetry currents.

We are interested in planar black hole solutions in AdS and so we
will consider the topology of the AdS timelike boundary to be
$R^3\times R^+$ --- the intrinsic coordinates of the boundary
are denoted by $x$ and the radial coordinate by $r$.
A concrete prescription to do computations on the Euclidean section
was given in \cite{Witten:1998qj}. That is, the on-shell bulk partition function with
the boundary conditions $\Phi(x,r)|_\infty=\Phi_0(x)$ is the generating
functional of the boundary QFT correlation functions:
\ben
Z_{SUGRA}[\Phi_0(x)]=\int\left<e^{\int d^4x\Phi_0(x)\mathcal{O}(x)} \right>_{QFT}.
\een
We work in the saddle point approximation and so the generating function
of QFT is determined by the on-shell supergravity action,
$W_{QFT}[\Phi_0]=-S_{onshell}[\Phi_0]$. To compute the two-point function
of a particular operator in the field theory, we should first solve the
linearized equation of the corresponding supergravity field with the
appropriate boundary conditions, then evaluate the on-shell sugra action,
and, finally, compute the second functional derivative with respect to
the source:
\ben
<O(x)O(y)>=-\frac{\delta^2S_{onshell}}{\delta\Phi_0(x)\delta\Phi_0(y)}\Big|_{\Phi_0=0}.
\een

We are interested in a bulk gravity theory with an electromagnetic
field non-minimally coupled to a scalar. In general there are two
types of perturbations of AdS: those that modify the bulk but
preserve the AdS asymptotics (e.g., a black hole is interpreted
as a finite temperature state in dual QFT) and those that do not
modify the bulk and act as sources for operators in the dual QFT.
The AdS metric (graviton) couples to the stress tensor and the
gauge field $A_\mu$ couples to the R-charged current $J^{\mu}$
of the field theory.

To obtain the viscosity (conductivity) we should compute
the two-point functions of the stress tensor (current). Due to
the coupling of the gauge field with a scalar, there will be a
non-trivial $3$-point function mixing two currents with a scalar
operator --- in this work, we are not interested in this correlation
function.

The bulk solution is a geometry with a horizon (at zero or finite
temperature) and to get the information about the hydrodynamics one should compute
the retarded Green functions directly on the Lorentzian section. One
reason is that, on the Euclidean section, there is a discrete set
of frequencies, the Matsubara frequencies ($\omega=2\pi inT$ for bosons
and $\omega=i\pi T(2n+1)/2$ for fermions). However,
since we are interested in a small $\omega$ in the hydrodynamic
approximation, an analytic continuation from the Euclidean section
where there is a minimum frequency ($\omega_m=2\pi iT$) is not useful.

The general prescription for real-time retarded Green's functions
within AdS/CFT duality can be found in \cite{Son:2002sd, Herzog:2002pc}.
The existence of hydrodynamics modes is reflected by the existence
of the poles of the retarded correlators. However, to compute the viscosity
bound, we will use a different prescription, namely the
membrane paradigm proposal of \cite{liu}.

We would like to present now a discussion on the boundary
conditions for gauge fields. The bulk
gauge field (\ref{Fansatz}) is invariant under translations in the
boundary coordinates ($x$). The component tangent to the boundary of
the gauge field is constant at large $r$, which corresponds to a
magnetic field $B_z$ perpendicular on the $(xy)$-plane.

If there is an electric field, $E_x$, in $x$-direction, then the
effect of a magnetic field in $z$ direction induces a
current along $y$-direction.
For,
$A_t(x)\neq 0$ and $A_i=A_y(x)\neq 0$, the conserved
currents on the boundary are the charge density, $q$, and the current
density, $J$ --- they are defined as
\be
q=\frac{\delta S}{\delta A_t}\Big|_{boundary} \, , \,\,\,\,\,\,\,\,\,\,
J^i=\frac{\delta S}{\delta A_i}\Big|_{boundary}.
\ee
To obtain the boundary current, we have to impose the following boundary
conditions:
\be
\delta A_t |_{boundary}=0\, , \,\,\,\,\,\,\,\,\,\, \delta A_i|_{boundary} \,\,\,\mathrm{free}.
\ee

The condition of regularity of the gauge potential at the horizon requires
that $A_t$ should vanish (on the Euclidean section, the radius of the Euclidean
time shrinks to zero at the horizon). Therefore, we should add a gauge term,
$\Upsilon$ so that $A_t(r_h)-\Upsilon=0$. This term, which remains constant
at the boundary, plays the role of a chemical potential for the electric charge.

Let us end up this section with a comment about the Chern-Simons term. The
coupling of the gauge Chern-Simons term is proportional to the chiral anomaly
of the corresponding current in the dual field theory. The current anomaly is
given by a one-loop triangle Feynman diagram to the $3$-point function of
(R-)currents. Therefore, there also exists a {\it chiral} magnetic current
which is induced parallel to the applied magnetic field, $J^z\sim \zeta B \Upsilon$.

\section{Near horizon geometry}\label{nhsec}
Shear viscosity of the boundary fluid can be computed from the knowledge
of the near horizon physics only \cite{liu} --- in Section 4, we
will present a detailed analysis. This observation is very useful when the
bulk Lagrangian is very complicated. For example, for the model we are
interested in, gravity is coupled to various matter fields in a non-trivial
way and it is technically difficult to find an analytic solution of the system.

However, it is not a very complicated problem to find the near horizon geometry,
i.e. how the metric and other fields behave in the near horizon limit. The
reason is that, in principle, we do not have to solve any differential
equation to find the near horizon geometry. A study of the extremal near
horizon geometries is interesting in its
own because it provides information about the instabilities that may
appear in the theory.

The algorithm is as follows: first we obtain the field equations. Then,
we consider a suitable near horizon ansatz for different fields --- all
fields should be regular at the horizon. Substituting the ans\"{a}tze for
fields in the corresponding equations of motion, one can solve them
consistently order by order in $(r-r_h)$.

In this section, we will first find the near-horizon geometry of
the non-extremal black hole, which we will need to compute
the shear viscosity to entropy density ratio in Section 4. In
the extremal limit, we will see that, due to the attractor mechanism,
the near horizon geometry is universal regardless of the asymptotic
values of the scalars. We present a detailed analysis of the
branches of solutions with finite horizon area.

\subsection{Non-extremal case}

As in \cite{D'Hoker:2009bc}, we work with a coordinate system in which
the solution takes the canonical form at the horizon. That is, the field
strength $F_H$ and the metric $ds_H^2$ are
\ben
\label{nearh}
F_H & = & q \, dr\wedge dt + B \, dx \wedge dy -p \ dr\wedge dz ,
\nn \\
ds_H^2 & = & r_H^2(dx^2 + dy^2 + dz^2) ,
\een
where $q$ and $B$ are the charge density (of the black brane) and
the magnetic field at the horizon, respectively. In this way, the
gauge freedom is removed and the initial conditions are
\ben
U(r_h)=Z(r_h)=P(r_h)=0\,\,\, , \,\,\,\,\,\,\,\,\,  V(r_h)=W(r_h)=\ln(r_h).
\een
A similar analysis (and numerical solutions) in the presence of
the Gauss-Bonnet term but without the Chern-Simons term was presented
in \cite{Astefanesei:2008wz}.

The generic solutions have a non-degenerate horizon. Near the event
horizon, they admit a power series expansion of the form (using the
definition of the temperature (\ref{temp}) in the expression of $U$):
\ben
\label{nonextremal}
U(r)&=& 4 \pi T (r-r_h)+ u_2 (r-r_h)^2 + \cdots , \nn\\
V(r)&=& \ln(r_h)+ v_1 (r-r_h)+ v_2 (r-r_h)^2 + \cdots ,\nn\\
W(r)&=& \ln(r_h)+ w_1 (r-r_h)+ w_2 (r-r_h)^2 + \cdots ,\nn\\
Z(r)&=& z_1 (r-r_h)+ z_2 (r-r_h)^2 + \cdots ,\nn\\
E(r)&=& q (r-r_h)+ q_1 (r-r_h)^2 + q_2 (r-r_h)^3 +\cdots ,\nn\\
P(r)&=& p (r-r_h) + p_1 (r-r_h)^2 + \cdots ,\nn\\
\varphi(r)&=& \varphi_h + \varphi_1 (r-r_h)+\varphi_2 (r-r_h)^2 + \cdots
\een
It is important to emphasize that, what is generally called
near horizon geometry for a non-extremal black hole is just a
truncation of the above series expansion. To compute the shear
viscosity, though, we need also some data at the order $(r-r_h)^2$.

Another observation is that, in principle, one can use a boost
transformation in $z$ direction to set $p=0$ (see \cite{D'Hoker:2009bc}).
However, the boost transformation is singular at some point
outside the black hole horizon. In our analysis we keep the value of
$p$ non-zero and determine
it in terms of other horizon data. We will see in Section 4 that
the expressions for the entropy, shear viscosity, and
their ratio remain unchanged if we set the horizon value of $P'(r)$
to be zero (in other words, they do not depend of $p$). This is
expected due to the fact that the physical quantities should
be invariant under the boost transformations.

By substituting the ansatz (\ref{nonextremal}) in the field equations,
we get the following expressions for the coefficients at the order
$(r-r_h)$:\footnote{We obtain the results as functions of the coefficients
($T, \ q, \ B, \ \varphi_h$, $r_h$, and $z_1$) --- this will simplify the
computations of the shear viscosity.}
\ben
\label{AA}
v_1&=&-\frac{2 B^2 e^{\alpha  \varphi _h}+r_h^4 \left(q^2 e^{\alpha
   \varphi _h}-6\right)}{6 \pi  T r_h^4},
\nn \\
w_1&=&\frac{4 B^2 e^{\alpha  \varphi _h}-r_h^4 \left(4 q^2 e^{\alpha
   \varphi _h}+3 z_1^2 r_h^2-24\right)}{24 \pi  T r_h^4},
\nn\\
p&=& \frac{q \left(2 B \zeta  e^{-\alpha  \varphi _h}+z_1
   r_h^3\right)}{4 \pi  T r_h},\nn \\
\varphi_1 &=& \frac{\alpha  e^{\alpha  \varphi _h} \left(q^2
   r_h^4-B^2\right)}{4 \pi  T r_h^4},
\nn \\
q_1&=& {e^{-2 \alpha  \varphi _h} \over 16 \pi
   T r_h^4}\bigg( 2 B^2 [\left(\alpha
   ^2+2\right) e^{3 \alpha  \varphi _h} +4 \zeta ^2]\nn \\
&& -r_h^4
   e^{2 \alpha  \varphi _h} [2 q^2 \left(\alpha ^2-2\right)
   e^{\alpha  \varphi _h}+z_1^2 r_h^2+24]\bigg).
\een

However, in higher derivative gravity theories the only
coefficient at the order $(r-r_h)^2$, which we need for viscosity
bound computation, is $u_2$. But, for completeness, we
present the expressions of all the other coefficients
that appear at order $(r-r_h)^2$, in Appendix $B$.

A non-extremal charged scalar black hole is characterized by
four independent parameters: the mass, electric charge, magnetic
field, and also the value of the scalar at the horizon, $\varphi_h$.
In this case, the horizon radius (and so the entropy) and the
horizon value of the scalar depend of the asymptotic boundary
data ($\varphi_\infty$). We will see in the next subsection that
this is in contrast with the extremal case for which we obtain
an attractor behaviour at the horizon.

At first sight, it may seem surprising that the data
(\ref{AA}, \ref{BB}) we need to compute the entropy and
shear viscosity depend also on $z_1$, a
coefficient that we do not compute explicitly. However, we will see
in Section 4 that the final values of
the physical quantities depend in fact just on four independent
parameters, namely $(q,B,r_h,\varphi_h)$ --- we `trade' the mass
for the horizon radius and so the independent parameters that
completely characterize the black hole are the ones mentioned above.

\subsection{Extremal limit and attractor mechanism}\label{extremal}
It is by now well known that the extremal black holes in theories of
Einstein gravity with scalar fields non-trivially coupled with
abelian gauge fields have an enhanced symmetry of the near horizon
geometry \cite{nearhorizon}. That is, the near horizon geometry contains
an $AdS_2$ spacetime.

The attractor mechanism \cite{Ferrara:1995ih} is also valid for non-BPS extremal
black holes \cite{Sen:2005wa, Sen:2005iz, Goldstein:2005hq} \footnote{A
generalization to stationary non-susy black holes was given in \cite{Astefanesei:2006dd}.
More details on the non-supersymmetric attractors can be found in the
nice reviews \cite{attractorreviews}.} --- a qualitative
explanation is that, due to the infinite long throat of $AdS_2$, the
near horizon geometry has no memory of asymptotic data. Consequently, the
near horizon geometry is universal and the entropy does not depend on the asymptotic
values of the scalars.

In this section we present a careful analysis of the extremal near
horizon geometry by using directly the equations of motion and also
the entropy function formalism. The advantage of the latter is that
we can compute the physical charges and obtain {\it analytic} expressions
for the entropy density. In this way, we can confirm the numerical analysis
of \cite{D'Hoker:2010rz}.

The (non-supersymmetric) attractor mechanism in AdS and its embedding
in string theory were studied in \cite{Astefanesei:2007vh}. In the rest
of this section we follow closely \cite{Astefanesei:2007vh, Astefanesei:2008wz},
though we should keep in mind that our goal is not to investigate just
the flow of the entropy but also the flow of some hydrodynamic transport
coefficients, and we also consider a non-trivial Chern-Simons term.

\subsubsection{Attractor mechanism}

In $AdS$ spacetime, the $BPS$ condition is different than the extremal
limit. For static black holes, the BPS limit is a naked singularity, but
in the extremal limit the entropy can remain finite.

Unlike the non-extremal case, in the extremal case the near horizon
geometry is an {\it exact} solution of the equations of motion and not
a truncation in a Taylor expansion. Let us consider the most general
ansatz for the near horizon geometry:
\ben
ds^2 &=& L\lb {dr^2 \over r^2} -r^2 dt^2  \rb + \upsilon_1 \left ( dx^2 + dy^2 \right  )
+ \upsilon_2 \left ( dz + z_1 r dt \right )^2\ , \nn\\
F&=& q dr\wedge dt + B dx\wedge dy , \nn \\
\varphi(r) &= &\varphi_h .
\een
Another important difference with the non-extremal case is that, due
to the enhanced symmetry of the near horizon geometry, $p$ vanishes
(the $z$ component of the gauge field is constant).
As a consistency check, it can be shown that with this ansatz, in
the absence of the scalar field, we obtain the results of
\cite{D'Hoker:2009bc}.\footnote{We emphasize again that
the authors of \cite{D'Hoker:2009bc} have shifted the
radial coordinate, $r$, to set the horizon at $r_h=1$.}

Let us consider now in more detail the non-trivial case when the scalar
is turned on. The first Maxwell equation can be easily integrated and
we obtain
\be \label{phycharge2}
Q_p={\upsilon_1^2 \upsilon_2 q^2 e^{2\alpha \varphi_h} \over L^2} ,
\ee
where the integration constant is related to the physical
charge.\footnote{$Q_p$ includes the contribution from constant
$z$ component of gauge field.}

We use the following system of independent equations:\footnote{The
first one is the $(rr)$-component of Einstein equation, the second
equation is a combination of $(rr)$- and $(tt)$-components, the third one
is the $(xx)$-component, the fourth one is the $(tz)$-component, the next
one is a combination of $(xx)$- and $(zz)$-components, and the last two
are the remaining non-trivial Maxwell equation and the scalar equation,
respectively.}
\be
\label{system}
4 B^2 L^2 \upsilon_2 e^{2 \alpha  \varphi _h}+\upsilon_1^2 \upsilon_2 e^{\alpha
   \varphi _h} \left(24 L^2-6 L+3 \upsilon_2 z_1^2\right)+8 L^2 Q_p^2=0,\nn
\ee
\be
4 B^2 L^2 \upsilon_2 e^{2 \alpha  \varphi _h}-4 L^2 \left(Q_p^2-6 \upsilon_1^2
   \upsilon_2 e^{\alpha  \varphi _h}\right)-3 \upsilon_1^2 \upsilon_2^2 z_1^2
   e^{\alpha  \varphi _h}=0,
\nn
\ee
\be
-2 B^2 e^{\alpha  \varphi _h}-\frac{Q_p^2 e^{-\alpha  \varphi
   _h}}{\upsilon_2}+6 \upsilon_1^2=0,
\nn
\ee
\be
\upsilon_2 \left(\frac{4 B^2 e^{\alpha  \varphi
   _h}}{\upsilon_1^2}+24\right)-\frac{4 Q_p^2 e^{-\alpha  \varphi
   _h}}{\upsilon_1^2}-\frac{3 \upsilon_2^2 z_1^2}{L^2}=0,
\ee
\be
\frac{2 B^2 \upsilon_2 e^{\alpha  \varphi _h}}{\upsilon_1}-\frac{\upsilon_1 \upsilon_2^2
   z_1^2}{2 L^2}=0,
\nn
\ee
\be
Q_p\lb \frac{8 B L \zeta  e^{-\alpha  \varphi _h}}{\upsilon_1 \sqrt{\upsilon_2}}+4
   z_1\rb=0,
\nn
\ee
\be
\alpha \lb B^2 L^2-\frac{L^2 Q_p^2 e^{-2 \alpha  \varphi _h}}{\upsilon_2}\rb=0\nn.
\ee

As we already explained above, we expect that the near horizon geometry
to be completely fixed by the charges. By solving the system of
equations (\ref{system}), we obtain the following expressions for $AdS_2$ radius,
size of transverse space, strength of fibration, and horizon value of the scalar:
\be
L={1\over 12} \, , \,\,\,\,\, \upsilon_1={B \zeta^{1/3}\over \sqrt{2}}
\, , \,\,\,\,\, \upsilon_2=\lb {Q_p\over B \zeta^{2/3}} \rb^2 \, , \,\,\,\,\,
z_1 = -{B \zeta^{2/3}\over 3 \sqrt{2} Q_p}\, , \,\,\,\,\,
\varphi_h = {2 \ln(\zeta)\over 3 \alpha}.
\ee
The radius of $AdS_2$ is the same as the radius of $AdS_5$. This is due to the
fact that we work with planar black holes --- for extremal spherical black holes
in $AdS_5$, the radius of $AdS_2$ does not match the radius of $AdS_5$
\cite{Astefanesei:2007vh}.

One important observation is that the horizon value of the scalar is
fixed by the Chern-Simons coupling. We will reobtain this result by
using the entropy function.

For a regular extremal solution whose entropy does not vanish,
there is a large ground state degeneracy. However, an important
question is then, if the entropy is `stable' against changes of
the boundary conditions for fields in the bulk. In \cite{D'Hoker:2009bc},
the authors argue that the near horizon
geometry ($AdS_2\times R^3$) of the purely electrically charged
brane is unstable under the addition of a magnetic field.

In the rest of this section we argue that, in the presence of
the scalar and electromagnetic field, the entropy does not change
under general perturbations of the scalar. A similar discussion for a
$5$-dimensional theory without the
Chern-Simons term can be found in \cite{Astefanesei:2007vh,
Astefanesei:2008wz}.\footnote{Recently, the authors of
\cite{Goldstein:2009cv} have used the same kind of arguments to
discuss charged scalar black holes in $AdS_4$.}

The arguments are as follows. Due to the non-trivial coupling
between the scalar and gauge fields, there is an `effective
potential' for the scalar \cite{Goldstein:2005hq}. If the
effective potential has a stable minimum at the horizon, the
solution is regular and the entropy is determined completely
by the charge parameters (if there are flat directions, not
all moduli are stabilized, but the entropy does not depend
of the flat directions).

We can see that, without a gauge field, the theory is scaling
invariant and so the scalar field is always massless. By turning
on just one gauge field (electric or magnetic) the symmetry is
broken and one expects that
the solution is not regular. The reason is that, even if
an effective potential is generated, the potential has no
stable minimum.

However, by turning on a second gauge field (in our case, we
consider an electromagnetic field) we expect to obtain regular
solutions. The effective potential can have stable minima and
the near horizon geometry is universal.

One important observation, though, is that the ansatz of our
solution is different than the one of \cite{Goldstein:2005hq} --- we
have non-trivial terms $(dzdt)$ in the metric. A concrete expression
for the effective potential in this case is not known. However, in
the near horizon limit the effective potential method should be
equivalent with the entropy function method of Sen. In what follows,
we will use the attractor equations to find the near horizon
geometry (and so the entropy density) of the extremal solutions.

\subsubsection{Entropy function}
By using the entropy function formalism we can
explicitly compute the physical charges and obtain an analytic
expression for the entropy density. Since our theory contains
the gauge Chern-Simons term, which is not gauge invariant, we
have to Kaluza-Klein (KK) reduce our action to four dimensions.

Let us consider first the case with the scalar turned off. We consider
the $z$ direction compactified on a circle of
radius $\beta$ and a general KK ansatz
\ben
g_{\alpha\beta}dx^{\alpha}dx^{\beta}&=&G_{ab}dx^adx^b+G_{AB}(dy^A+\bar{A}_a^Adx^a)(dy^B+\bar{A}_a^Bdx^a),\nn\\
A^{(5)}&=&A_\mu^{(5)}dx^\mu = A_a^{(4)}dx^a + C_B(x^a) (dy^B + \bar{A}^B_a dx^a) ,
\een
where ${a,b}$ are $4$D indices and ${A,B}$ are compact indices
($z$ in our case). With this notation we have splitted the
coordinates as $x^\mu=(x^a, y^A)$ and so $A_\mu^{(5)}$ is the
$5$-dimensional gauge potential, $A_a^{(4)}$ is the $4$-dimensional
gauge potential, and $\bar{A}^B_a$ is the KK gauge potential.

We use the following results of the dimensional reduction
\cite{schwarz} (there is no dependence of KK coordinates):
\ben
\sqrt{-g}&=&\sqrt{-G}\sqrt{\det(G_{AB})}\,\,\, , \,\,\,\,\,\,\,\,\,\,\,
F^{(5)2} = F^{(5)\mu\nu} F^{(5)}_{\mu\nu} = F^{(5)ab} F^{(5)}_{ab} + 2F^{(5)aB} F^{(5)}_{aB}\, ,\nn\\
R_5&=&R_4-\frac{1}{4}G^{ac}G^{bd}G_{AB}\bar{F}^{A}_{ab}\bar{F}^{B}_{cd}+\frac{1}{4}
\partial_aG_{AB}\partial^aG^{AB}-\frac{1}{4}G^{AB}\partial_{a}G_{AB}G^{CD}\partial^{a}G_{CD}-\nn\\
& & -\partial_a(G_{AB}\partial^aG_{AB}).
\een
Now, we are ready to write down the ansatz we are interested in --- the metric, gauge field, and
relations between the $5$-dimensional gauge potential, KK gauge potential, and $4$-dimensional
gauge potential are
\ben
\label{AnsatZ}
ds^2&=& L (-\frac{dt^2}{r^2}+r^2dr^2) + \upsilon_1 (dx^2+  dy^2) + \upsilon_2(dz + z_1 r dt)^2, \nn \\
A_{\mu}^{(5)}dx^{\mu} &=& \vartheta r dt - \frac{B}{2} y dx + \frac{B}{2} x dy - p_1 (dz+z_1 r dt),\nn \\
F_{rt}^{(5)}&=& q = \vartheta - p_1 z_1\, , \,\,\, F^{(5)}_{xy}=F^{(4)}_{xy}=B\, , \,\,\,  F_{rt}^{(4)} = \vartheta \, , \,\,\,
\bar{F}_{rt}^{z} = z_1.
\een
The on-shell action and entropy function are
\ben
S&=& \frac{{\cal A}_{xy} \beta} {16 \pi G_5} \left[L \sqrt{\upsilon_1^2 \upsilon_2}
\left(-\frac{2 B^2}{\upsilon_1^2}+\frac{2 (\vartheta-p_1 z_1)^2}{L^2}
+\frac{\upsilon_2 z_1^2-4 L}{2 L^2}+12\right)+8 B p_1 \zeta  \lb\frac{p_1 z_1}{2} -\vartheta\rb\right], \nn \\
{\cal E} &=& 2 \pi \beta {\cal A}_{xy} \lb Q\  \vartheta + \Theta\  z_1- {S\over \beta {\cal A}_{xy}}\rb ,
\een
where $Q$ is the $4$-dimensional physical charge, $\Theta$ is the physical
charge associated to $KK$ gauge field, and ${\cal A}_{xy}=\int dx dy$.

The equations of motion in the near horizon limit are
\ben
\label{nnn}
-\frac{\beta \pi \sqrt{\upsilon_2}}{L \upsilon_1^2} \left[4 B^2 L^2 + \upsilon_1^2 (-4 L + 24 L^2 + 4 q^2 + \upsilon_2 z_1^2)\right]&=&0 , \nn \\
\frac{\beta \pi}{L \upsilon_1 \sqrt{\upsilon_2}}\left[4 B^2 L^2 - \upsilon_1^2 (-4 L + 24 L^2 + 4 q^2 + 3 \upsilon_2 z_1^2)\right]&=&0 , \nn \\
\frac{\beta \pi \sqrt{\upsilon_2}}{L \upsilon_1^2}\left[4 B^2 L^2 + \upsilon_1^2 (-24 L^2 + 4 q^2 + \upsilon_2 z_1^2)\right] &=&0\\
\frac{2 \beta \pi}{L}\left[L \left(\Theta-4 B p_1^2 \zeta \right)+\upsilon_1 \sqrt{\upsilon_2} (4 p_1 q-\upsilon_2 z_1)\right]&=&0 , \nn \\
\frac{8 \beta \pi q}{L}\left[2 B L \zeta + \upsilon_1 \sqrt{\upsilon_2} z_1\right] &=&0 , \nn \\
2 \pi  \beta \left(8 B p_1 \zeta -\frac{4 \upsilon_1 \sqrt{\upsilon_2} q}{L}+Q\right)&=& 0.\nn
\een
First, we solve the last equation to obtain the $5$-dimensional electric
potential. Then, we use this result in all the other equations, and from
the forth equation
we obtain the following expression for the KK potential:\footnote{We note that by
combining the first equations we get $L=1/12$.}
\be
z_1 ={L (p_1 Q + \Theta + 4 B p_1^2 \zeta) \over \upsilon_1 \upsilon_2^{3/2} }.
\ee
Next, we compute $p_1$ from the sum of the first two equations --- we obtain
a 4th order equation with the following four solutions:
\ben
p1 = - {Q + \sqrt{Q^2 - 16 B (2 B \upsilon_2 + \Theta) \zeta} \over 8 B \zeta}\,\, , \,\,\,
p1 = {-Q + \sqrt{Q^2 - 16 B (2 B \upsilon_2 + \Theta) \zeta} \over 8 B \zeta} \nn \\
p1 = {-Q + \sqrt{Q^2 + 16 B (2 B \upsilon_2 - \Theta) \zeta} \over 8 B \zeta} \,\, , \,\,\,
p1 = -{Q + \sqrt{Q^2 + 16 B (2 B \upsilon_2 - \Theta) \zeta} \over 8 B \zeta}. \nn
\een
In the absence of the magnetic field, we expect a trivial fibration for our geometry:
we can see that the physical solutions are the second and third ones because $p_1$ is finite
in the limit $B\rightarrow 0$. Therefore, depending on $p_1$ solution,
we have two distinct families of solutions.
To compute the other near horizon parameters, we proceed as
follows: by subtracting the third equation from the first one we solve for $\upsilon_2$,
and get
\be
\upsilon_2={L (Q^2 - 16 B \Theta \zeta)\over 8 [\upsilon_1^2 + 4 B^2 L (-1 \pm \zeta)]}.
\ee
The $+/-$ solution corresponds to the second and third solution for $p_1$, respectively.

The $5$th equation, which is very important for understanding the solutions, and
the entropy are
\be\label{extcond}
B(1\pm \zeta) \frac{\sqrt{3\upsilon_1^2-B^2}}{36 \sqrt{2}\upsilon_1}=0\,\,\,\,\,\, , \,\,\,\,\,
S=\beta \upsilon_1 \pi \sqrt{\frac{Q^2-16 B \Theta \zeta}{6 \upsilon_1^2 + 2 B^2(-1\pm \zeta)}}.
\ee

It is now clear that, from the equation (\ref{extcond}), there are three
distinct branches:\footnote{We rewrite $q$ in terms of the physical parameters
$q=\frac{\sqrt{3\upsilon_1^2-B^2}}{6 \sqrt{2}\upsilon_1}$ to obtain the constraint
between $\upsilon_1$ and $B$ for $q=0$.}
\be
B=0, \ \ \ q=0\,(\upsilon_1^2=B^2/3), \ \ \ \zeta = \pm1.
\ee
From (\ref{extcond}), we can also read off the corresponding entropy for all three
branches:\\

\noindent
\underline{{\it Branch 1}: $B=0$}\\

This branch corresponds to the usual Reissner-Nordstr\"om solution and its entropy is
\ben
S&=& {\beta {\cal A}_{xy} \pi Q \over \sqrt{6}}.
\een

\noindent
\underline{{\it Branch 2}: $q=0$ ($B/ \upsilon_1= \sqrt{3}$)}\\

We compute all near horizon parameters in terms of physical charges. We obtain
$p_1=-{Q \over 8 B \zeta}$ and also find two possible solutions for $z_1$ and $\upsilon_2$:
\ben
\rm{set 1}:\ \ \ \ z_1= - {2 \sqrt{2}B \over \sqrt{3}\sqrt{{Q^2\over \zeta} - 16 B \Theta}} \ , \ \ \ \
\upsilon_2 = {1\over 32 B^2} \lb {Q^2\over \zeta} - 16 B \Theta\rb,
\een
\hspace{7cm}  or
\ben
\rm{set 2}:\ \ \ \ z_1=  {2 \sqrt{2}B \over \sqrt{3}\sqrt{16 B \Theta -{Q^2\over \zeta}}} \ , \ \ \ \ \
\upsilon_2 = {1\over 32 B^2} \lb16 B \Theta -{Q^2\over \zeta} \rb .
\een
The corresponding entropies  are
\be
S_1 = {\beta {\cal A}_{xy}\pi \over \sqrt{6}} \sqrt{{Q^2\over \zeta}-16 B \Theta}, \,\,\,,\,\,\,\,\,\,
S_2 = {\beta {\cal A}_{xy}\pi \over \sqrt{6}}\sqrt{16 B \Theta-{Q^2\over \zeta}}.
\ee

\noindent
\underline{{\it Branch 3}: $\zeta = \pm 1$} \\

In this case, one of the near horizon equations of motion, in fact,
is used to get a fixed value for $\zeta$. Therefore, we can not compute
all the parameters, one of them can not be fixed. However, as expected,
the entropy depends only on the physical charges:
\ben
S&=&\beta {\cal A}_{xy} \pi \sqrt{\frac{Q^2\mp 16 B \Theta }{6}}\ \ \ \ \ when \ \ \zeta=\pm1.
\een

Now, we are ready to compare our analytic results with the numerical
results of \cite{D'Hoker:2010rz}. We just saw that with the entropy
function formalism we get a similar condition as the one of
\cite{D'Hoker:2009bc, D'Hoker:2010rz}, namely $qB(1\pm\zeta)=0$.
Therefore, we also obtain solutions for $qB=0$ or $\zeta=\pm1$.
In the first case, when $\zeta\neq \pm1$, there is a regular
solution (when $B=0$, branch 1) with the expected near horizon geometry
($AdS_2\times R^3$).

For the case $q=0$ (branch 2), depending on the sign of $\zeta$,
we have different situations. Let us assume that the physical $KK$ charge is
$\Theta > 0$. When $\zeta>0$ (but $\neq 1$), the positivity of
$\upsilon_2$ (which is also reality of $z_1$) implies that
$B$ has a critical value $B_c={Q^2\over 16 \Theta \zeta}$ (solution set 1).
Once $B$ crosses that critical value, the solution set 1 becomes singular and
the solution set 2 takes over.  In this case, for a generic value of the
magnetic field, the near-horizon geometry is always $AdS_3 \times R^2$
(see, also, \cite{D'Hoker:2009bc}). For $\zeta<0$ (and $\neq -1$),
the solution set 1 does not exist for any positive value of the magnetic field.
In this case, the only possible solution is set 2, which is a regular
(finite horizon area) solution for any positive value of $B$.
We would also like to point out that,
when the Chern-Simons coupling vanishes, it seems there is no
finite area solution in the case $q=0$ and non-zero magnetic field.

In the second case, $\zeta=\pm1$,
there is a family of solutions for which
$2(B/v_1)^2 +(q/L)^2-6=0$ (this constraint is obtained by solving
the third and fifth equation in (\ref{nnn}) and matches
the condition of \cite{D'Hoker:2009bc} where the horizon radius
is rescaled so that $v_1=1$). In this case for $\zeta=1$ there exists
a critical magnetic field $B_c={Q^2\over 16 \Theta}$ above which the
solution becomes a naked singularity. For $\zeta =-1$ the solution exists
for any value of $B$.

The near horizon geometry smoothly interpolates between
$AdS_2\times R^3$ (for $B=0$) and $AdS_3\times R^2$
(for $q=0$) --- in between, the near horizon geometry
is in fact an warped $AdS_3\times R^2$.\footnote{The
$AdS_3$ spacetime is a special fibration of $S^1$ over
$AdS_2$ and the case $q=0$ corresponds to this special
fibration.}


Let us know discuss the case when the modulus is turned on. The
analysis is very similar with our computations above --- we
present the details in Appendix C. Unlike in the previous section
where we have solved the equations of motion in the near horizon geometry,
by using the entropy function we can compute the physical charges
in four dimensions. In five dimensions, due to the existence of
the gauge Chern-Simons term, one has to be careful about the definition of
the physical charge --- we will present a detailed analysis of
this aspect in the forthcoming work \cite{DNS2}.

By using the attractor equations from Appendix C, we obtain the
same horizon value of the modulus
\be
\varphi_h= {2 \over 3\alpha} \ln{|\zeta|}
\ee
as in the previous subsection. In general, due to the attractor mechanism,
in the extremal case the horizon values of the moduli are fixed by the
charges. Interestingly enough, this is not the case here: the modulus is
fixed by the Chern-Simons coupling alone. We will present in Discussion
section an argument of why this is the case.

Another observation is that the physical magnetic field is also fixed
by the Chern-Simons coupling:
\be
{B\over \upsilon_1}=\pm\, \sqrt{2}\, e^{-\alpha \varphi_h/2}
\ee
or, equivalently, the parameter $v_1$ is fixed only by the magnetic
field parameter and the Chern-Simons coupling and does not depend
on the electric field. This is due to the fact that we can rescale
$x$ and $y$ coordinates in our solution to make $v_1=1$, and this
rescaling affects just the magnetic field --- that can be easily
seen from our ansatz for the metric and gauge potential (\ref{AnsatZ}).

In Appendix C, we present a detailed computation of the entropy --- the
result is
\be
S= \beta {\cal A}_{xy} \pi \sqrt{{Q^2 -16 B \Theta \zeta \over 6 |\zeta|^{2/3}}}.
\ee

It is not our goal to make a detailed analysis with
the scalars turned on here (for different
$\alpha$ and `anisotropic near horizon geometries'), but
we observe that, when just the electric or magnetic field is
turned on, the solution is a naked singularity. This
is similar with our discussion at the end of the previous subsection,
when without the Chern-Simons term, we have argued that the
effective potential does not have a minimum and so there is
no regular solution \cite{Astefanesei:2007vh, Astefanesei:2008wz}.
We do not know the form of the effective potential when the
Chern-Simons term is present, but
that can be easily seen from the equation of motion of the scalar
(the last equation in (\ref{eoms}) ). Consequently, there is just
one branch of finite area solutions when the modulus is turned on.

As in the case with the scalar turned off, there also exists a critical
value of the magnetic field for which the entropy shrinks to zero
(for $\zeta>0$). So,
the geometry is stable against scalar fluctuations but not against
the magnetic field. However, this interpretation (and also the interpretations
of \cite{D'Hoker:2009bc, D'Hoker:2010rz}) should be taken with caution. It
is expected that the higher derivative corrections will `dress'
the singularity with a horizon and so a more general analysis is
important in this context.

\section{Shear Viscosity to Entropy Density Ratio}
\label{etabys}

For black objects with translation invariant horizon, for example
black brane geometry, one can also discuss the hydrodynamics --- long
wave length deviation (low frequency fluctuation) from thermal equilibrium.

In addition to the thermodynamic quantities the black brane is also
characterized by the hydrodynamic parameters like viscosity, diffusion
constant, electrical conductivity, {\it et cetera}. The black $D3$-brane geometry
with low energy fluctuations (i.e. with hydrodynamic behaviour) is dual
to a finite temperature gauge theory plasma living on boundary
with hydrodynamic fluctuations.

In this section we will study the low frequency behaviour of boundary
plasma dual to the gravity model presented in Section 2. To do this, we use
the membrane paradigm proposal of \cite{liu} and its generalization to
higher derivative gravity theories \cite{ns1}.

\subsection{Membrane paradigm and viscosity bound}
\label{membrane}

To an external observer, the black hole physics appears to be equivalent
to the physics of a dynamical fluid membrane \cite{thorne, pw1}. In other words,
the black hole is equivalent with a set of surface charges and currents
at the stretched horizon.\footnote{The stretched horizon is a timelike
surface just outside the true horizon. We emphasize that
the charges and currents are fictitious in the sense that
a falling observer through the stretched horizon does not detect any
surface sources. However, for an external observer, their presence is
consistent with all external field configurations.}

In the Lorentzian prescription of \cite{Son:2002sd} for computing
the Green functions, one should impose in-falling boundary conditions
for the fields at the horizon. However, when computing the two-point
functions, just the contribution from the boundary has to be taken.
An explanation of why the surface terms coming from the horizon must
be dropped was given in \cite{Herzog:2002pc} (see, also,
\cite{Maldacena:2001kr}): the boundary conditions should be defined
in Kruskal coordinates and one has to work with the full analytic extension
of the black hole.

An important observation, which relates the work of \cite{Son:2002sd} with
the membrane paradigm, was made in \cite{liu} (see, also, \cite{sonsh, cai1}).
That is, the response functions in AdS/CFT
are similar with the membrane response when the membrane is `pushed' at the
boundary. In this way, the membrane paradigm physical quantities
are in fact concrete gauge theory observables.\footnote{The infalling
boundary conditions at the horizons are the regularity conditions in the
membrane paradigm: near the horizon the fields depend trough a non-singular
combination of $r$ and $t$, namely $dv=dt+\sqrt{g_{rr}/g_{tt}}dr$.}

Interestingly, it was observed in \cite{nsnh} that only with
the knowledge of the near horizon geometry one can easily calculate
the shear viscosity of boundary fluid. One does not need to know the
full analytic solutions of Einstein equations --- this method is
especially useful in higher derivative AdS gravity theories.

As a warm-up exercise, let us consider first the model presented in
Section $\ref{model}$. We apply the method of \cite{liu} to compute
the shear viscosity coefficient of the boundary fluid for the action
(\ref{action1}).

Let us consider a metric perturbation of the form:
\be\label{petmet}
g_{xy}=g^{(0)}_{xy}+ h_{xy}(r,x)=g^{(0)}_{xy}[1+\ep \Phi(r,x)].
\ee

At first sight, it seems that the dual gravitational mode $(4.1)$ does
not generally decouple. Interestingly enough, the decoupling occurs
when the momentum vanishes and this is what we need for the computation
of the viscosity in the hydrodynamics limit --- we provide a detailed 
derivation of this claim in Appendix \ref{hxy}.\footnote{We would like 
to thank Alex Buchel for a discussion on this point.}

By plugging (\ref{petmet}) in the action and keeping the terms at order
$\epsilon^2$ (at the first order in $\epsilon$, we obtain
the equations of motion for gravitons), we get the following effective
action for the perturbation:\footnote{The terms that contain
the derivatives with respect to the spatial coordinates, $\vec{x}$,
combine in terms whose coefficient is proportional with
$\vec{p}^2$. Since we work in the hydrodynamic approximation
$\vec{p}=0$, these terms do not play any role in our analysis.}
\be \label{ghdacnphi1}
S=\nt \int {d^4 k \over (2 \pi)^4} dr \sum_{p,q=0}^{2} \cA_{p,q}(r,k)
\phi^{(p)}(r,-k) \phi^{(q)}(r,k).
\ee
Here, we use the Fourier transform to work in the momentum space $k=\{-\omega,\vec{k}\}$
\be \label{phifu}
\Phi(r,x) = \int {d^4k \over (2 \pi)^4} e^{i k.x} \phi(r,k) \, , \,\,\,\,\,\,\,\,\,
\phi(r,-k)=\phi^\ast(r,k)
\ee
and $\phi^{(p)}(r,k)$ denotes the $p^{th}$ derivative of the field $\ph$
with respect to $r$ ($p+q\leq 2$).

Next, we integrate by parts to obtain the bulk action for the graviton
in the following form (up to some total derivative terms):
\be \label{ghdacnphi2}
S=\nt \int {d^4 k \over (2 \pi)^4} dr [{\cal A}_1(r,k) \phi'(r,k)\phi'(r,-k) +{\cal A}_0(r,k) \phi(r,k)\phi(r,-k)] ,
\ee
where
\be
{\cal A}_1(r,k)=-{1\over 2}e^{2V(r)+W(r)} U(r), \ \ \ \ \ {\cal A}_0(r,k)={e^{2V(r)+W(r)}\omega^2 \over 2U(r)}.
\ee
At this point, it is important to emphasize that there are many
total derivatives in this action that do not affect the equations
of motion for the graviton. For the computation of the
{\it imaginary part} of the two-point function, the coefficient of
the term $\phi'\phi'$ in the bulk action is important. The other
total derivatives in the bulk action and the Gibbons-Hawking surface
term contribution exactly cancel on the boundary \cite{cai1}. It was
also shown \cite{cai1} (it is straightforwardly to check it also
in our case) that, in the case of Einstein gravity, the ratio of
viscosity and entropy density is not affected when the matter fields
are minimally coupled. The effective coupling \cite{Brustein:2008cg, cai1} is
\be
K_{eff}={1\over 16 \pi G_5}{{\cal A}_1(r,k) \over \sqrt{-g}g^{rr}}= -{1\over 32 \pi G_5}
\ee
and so the viscosity coefficient of the boundary fluid stress tensor is
\be
\eta = e^{2V(r_h)+W(r_h)}(-2 K_{eff}(r_h)).
\ee

In this case, the shear viscosity to entropy density ratio turns out to be universal, namely
\be
{\eta\over s}=\frac{1}{4 \pi }.
\ee

\subsection{Four derivative action}

Let us now consider the action (\ref{action1}) supplemented
with the most general four-derivative interactions \cite{mps}:
\begin{eqnarray}\label{hdaction}
S_{{\rm HD}}&=&S_{EM}+ \frac{\alpha'}{16 \pi G_5}\int d^5x\sqrt{-g} \bigg [c_1 R_{abcd}R^{abcd}
+c_2 R_{abcd}F^{ab}F^{cd} +c_3
(F^2)^2 \\ \nn
&& \ \ \ \ \ \ \ \ \ \ \ \ \ \ \ \ \ \ \ \ \ \ \ \ \ \ \ \ \ \ \ \ \ \ \ \
+ c_4\,F^4 +c_5\, \epsilon^{abcde} A_a R_{bcfg} R_{de}{}^{fg}\bigg ]\, .
 \nonumber
\end{eqnarray}

Since in supergravity actions the gauge kinetic terms couple to
various scalars, it will be interesting to understand the role
of the moduli in computing the viscosity bound. Unlike \cite{mps},
our action contains a scalar, $\varphi$, and the coefficients $c_i$
depend on the value of $\varphi$. This resembles the four-derivative
supergravity action \cite{Hanaki:2006pj}.

We treat the higher derivative terms perturbatively and apply the
method of \cite{ns1} to compute the shear viscosity coefficient of
the boundary fluid. However, to obtain the viscosity bound we also need the entropy
density. We start by using the Noether charge formalism of Wald
\cite{Wald:1993nt} (see, also, \cite{dg, Astefanesei:2008wz} for a discussion
in AdS) to compute the entropy density --- we will need just the
data in Section $3.1$ and Appendix B.

When we add higher derivative corrections to the action, the entropy
is no longer given by the area law --- instead, we use a general
formula proposed by Wald
\be
s=-2\pi \int_{{\cal
H}} {\partial L \over \partial R_{abcd}}\epsilon_{ab}\epsilon_{cd} ,
\label{eq:wald2}
\ee
where $\epsilon_{ab}$ is the binormal to the surface ${\cal H}$.

By using (\ref{eq:wald2}), we obtain the following expression for
the entropy density:
\be
\label{s}
s=\frac{r_h^3}{4G_5}+\frac{\alpha'}{4G_5}\,r_h^3\,[c_1(3z_1^2r_h^2-4u_2)-2c_2q^2]+{\cal}O(\alpha'^2).
\ee
We use the expression of $u_2$ given in Appendix B to rewrite this
expression as
\be
s=\frac{r_h^3}{4 G_5}-{\alpha' \ r_h^3\over G_5}\ltb {c_1\over 3}
\lb {5B^2\over r_h^4} e^{\alpha \varphi _h} + 7 q^2 e^{\alpha \varphi _h}-6\rb
+{c_2\over 2}q^2 \rtb +{\cal}O(\alpha'^2).
\ee
As expected, the entropy density depends on four independent parameters,
namely $(r_h,q,B,\varphi_h)$. Since in Wald formula only the four derivative
interactions that involve the curvature tensor are important, the entropy
only depends on $c_1$ and $c_2$ ($c_5$ does not appear because
the binormal has just $rt$ components and the contribution from this
term vanishes).

To compute the four derivative corrections to the shear viscosity
coefficient, we have to find the quadratic action for the
transverse graviton moving in the background spacetime. As in previous
section, we consider again the following metric perturbation
\be
g_{xy}=g^{(0)}_{xy}+ h_{xy}(r,x)=g^{(0)}_{xy}[1+\ep \Phi(r,x)] ,
\ee
where $\epsilon$ is an order counting parameter.

In the presence of a generic $n$ derivative term in the bulk Lagrangian,
the action (in momentum space) can be written as
\be \label{ghdacnphi}
S=\nt \int {d^4 k \over (2 \pi)^4} dr \sum_{p,q=0}^{n} \cA_{p,q}(r,k)
\phi^{(p)}(r,-k) \phi^{(q)}(r,k).
\ee

Here, $\phi^{(p)}(r,k)$ denotes the $p^{th}$ derivative of the field $\ph$
with respect to $r$ and $p+q\leq n$. The coefficients $\cA_{p,q}(r,k)$
depend, in general, on the coupling constant $\alpha'$.

It is important to emphasize that $\cA_{p,q}$ with $p+q \ge 3$
are proportional to $\alpha'$ and vanish in $\alpha' \ra 0$
limit \cite{ns1} --- these terms appear as an effect of higher
derivative terms in the action.

This action does not have the canonical form as in the two derivative
case. Therefore, it is not obvious how to generalize this approach for
higher derivative case. The proof given in \cite{liu} was based on the
canonical form of gravitons action. This problem was solved in \cite{ns1}
and we use this method to compute the shear viscosity coefficient.

Let us now write the effective action for the transverse graviton in
the canonical form with arbitrary coefficients:
\ben \label{acneff}
S_{eff}&=& {1 \over 16 \pi G_5}
\int {d\omega d^3{\vec k}\over (2 \pi)^4}dr \bigg[ {\cal B}_1(r,k) \cphp \php
+ {\cal B}_0(r,k) \ph \cph \bigg ] .\nn \\
\een

We demand that the equations of motion obtained from the action (\ref{acneff})
and original action (\ref{ghdacnphi}) match at order $\alpha'$. By comparing the
equations of motion for $\phi(r, k)$ from the two actions, we get the function
${\cal B}_0$ and ${\cal B}_1$.

Once we have the effective action for $\phi(r,k)$ in the canonical form, the effective
coupling can be easily read off:
\be\label{gkeff}
K_{eff}(r)=  {1\over 16 \pi G_5}\
{{\cal B}_1(r,k) \over
\sqrt{-g}
  g^{rr}} ,
\ee
where $g^{rr}$ is the $rr$-component of the inverse perturbed metric and
$\sqrt{-g}$ is the determinant of the perturbed metric. Therefore, the shear
viscosity coefficient is
\be
\eta = r_h^{{3}} [-2 K_{{eff}}(r=r_h)].
\ee
Evaluating the effective coupling in the near horizon we obtain the
shear viscosity coefficient
\ben
\eta&=&\frac{1}{16 \pi  G_5} \nn \\
 && +\frac{\alpha' \left(c_1 r_h^4 \left(8 q^2 e^{\alpha  \varphi _h}
+3 z_1^2 r_h^2-32 \pi  T v_1-36 \pi  T w_1-10 u_2+48\right)-B^2 \left(c_2-8 c_1
   e^{\alpha  \varphi _h}\right)\right)}{8 \pi  G_5 r_h^4}\nn\\
&& + {\cal O}(\alpha'^2)
\een
which can be rewritten as (we use the results in Appendix B)
\be
\eta=\frac{r_h^3}{16 \pi  G_5}-{\alpha' \ r_h^3\over 2 \pi G_5}  \ltb c_1 \lb q^2
+ {B^2\over r_h^4}\rb e^{\alpha \varphi _h} + {c_2\over 4}{B^2\over r_h^4}\rtb
+ {\cal O}(\alpha'^2).
\ee
The ratio of the shear viscosity and entropy density turns out to be
\be
{\eta\over s}=\frac{1}{4 \pi }+\frac{\alpha'}{\pi} \ltb {c_1\over 3}\lb (q^2
- \frac{B^2}{r_h^4}) e^{\alpha \varphi _h} -6\rb +{c_2\over 2}(q^2
- \frac{B^2}{r_h^4}) \rtb+ {\cal O}(\alpha'^2).
\ee
In $B\ra 0$ limit this result matches with \cite{mps}.\footnote{Note that our
`$q$' is different than `$q$' of \cite{mps}. In \cite{mps}, $q$ is the physical
charge (up to some normalization). In our case, the physical charge is
$\sim r_h^6 q$.}

Let us end up this section with a discussion of the extremal limit. In
the absence of the moduli, the extremality condition is
$2B^2  +r_h^4\lb q^2 -6\rb = 0$ and so the shear viscosity to entropy
density ratio becomes
\be
{\eta\over s}=\frac{1}{4 \pi }+\frac{\alpha'}{\pi} \ltb -c_1 {B^2\over r_h^4}
+ {3 c_2 \over 2} \lb 2- {B^2 \over r_h^4}\rb \rtb + {\cal O}(\alpha'^2).
\ee

Therefore, there is a drastic change when the magnetic field is turned on. That is,
unlike the electrically charged solution studied in \cite{mps}, the leading
correction of $\eta/s$ in the extremal limit depends on both, $c_1$ and $c_2$.
As expected, in $B\ra 0$ limit our result matches with the one of \cite{mps}.

In the presence of the scalar field, the extremal limit is more constrained.
The scalar is fixed by the Chern-Simons coupling only, and by using some
of the formulas derived in Section $3$, namely
\ben
2B^2 e^{\alpha \varphi_h} +r_h^4\lb q^2 e^{\alpha \varphi_h}-6\rb &=& 0 \nn\\
q^2 r_h^4 - B^2 & =& 0\nn \\
\varphi_h  - {2\over 3 \alpha} \ln |\zeta|&=& 0.
\een
we obtain the following expression for the shear viscosity to entropy
density ratio:
\be
{\eta\over s}=\frac{1}{4 \pi } \lb 1 - 8 c_1 \alpha' \rb + {\cal O}(\alpha'^2).
\ee
Interestingly enough, we observe that there is no dependence of the horizon value
of the modulus as in the non-extremal case. Also, the coefficient of $c_2$ vanishes
and so the dependence of $c_2$ drops out.

\section{Discussion}

The goal of this paper was two-fold. First, to extend the work of
\cite{D'Hoker:2009bc} by including scalar fields. Second, to compute
the low frequency transport coefficients of a $(3+1)$-dimensional
field theory whose gravity dual is an electrically and magnetically
charged planar black hole in AdS$_5$ Einstein-Maxwell theory with
a gauge Chern-Simons term.

Since, in the extremal case, we have used the entropy function method
to study the near horizon geometry, we were able to find analytic
expressions for the entropy density and the near horizon parameters.
In the absence of the scalar field we confirm the numerical results of
\cite{D'Hoker:2010rz}, in particular the existence of a critical value
of the magnetic field for which the entropy density vanishes. However,
we have found a larger class of solutions --- we have presented a detailed
discussion in Section $3$ and we do not want to repeat the details here.
However, we would like to point out that one of the special values of the
Chern-Simons coupling, $\zeta=1$, for which there exist finite horizon
area solutions with both, electric and magnetic, fields non-zero
corresponds to a special embedding of the gauge field
$U(1) \subset SU(4)_R$ \cite{D'Hoker:2009bc}.

In \cite{D'Hoker:2010rz}, it was proposed that $(3+1)$-dimensional gauge
theories holographically dual to $(4+1)$-dimensional
Einstein-Maxwell-Chern-Simons theory undergo a quantum phase transition
in the presence of a finite charge density and magnetic field. The
authors of \cite{D'Hoker:2010rz} also argue that a non-vanishing
entropy density is `exotic' from the point of view of CFTs that arise in the
AdS/CFT duality. By turning on the magnetic field, instabilities
can appear and this is consistent also with our analysis.

However, we believe that this result should be taken with caution.
It is well known that, in many situations, by adding higher derivative
corrections the singularities can be `dressed' by
horizons.\footnote{In \cite{D'Hoker:2010ij}, D'Hoker and Kraus have found a new
solution with a near horizon geometry that does not contain an
$AdS_2$. Based on the symmetries, one can argue that higher curvature
corrections do not change the near horizon form of the solution.}
Also, by turning on the moduli the geometries become more stable, though
there can appear other kind of instabilities, e.g. the ones associated
to the $AdS_2$ spacetime.

Since in string theory, typically, the gauge kinetic terms will couple
to various scalars \cite{Liu:2010sa}, we have considered a simple
extension of \cite{D'Hoker:2009bc} by considering an exponential scalar coupling.
In the presence of the moduli some important changes occur. First of
all, in the non-extremal case, the near horizon data depend on the boundary
value of the scalar. Therefore, the hydrodynamic properties of the dual
field theories with higher derivative terms will be also controlled
by the moduli.

Let us start with a discussion of the extremal case for
which there is a drastic change. That is, due to the attractor
mechanism, the near horizon geometry is universal and does not depend
of the asymptotic values of the moduli. The horizon area is finite
and so there is a large ground state degeneracy.
One important question is what is happening in the presence of scalars?
Is the ground state degeneracy still unstable  under inclusion of a
(large enough) magnetic field? Are there finite horizon area solutions
for any value of the Chern-Simons coupling when both (magnetic
and electric) fields are non-zero?

Before providing concrete answers to these questions, we would like
to emphasize that the attractor mechanism plays an important role in
classifying the bulk theories where the extremal entropy vanishes. A
detailed discussion of the attractor mechanism in AdS$_5$ and its
embedding in string theory was presented in \cite{Astefanesei:2007vh}.

The universality of dual theory, which means that the IR physics does
not depend upon the UV details, becomes in the holographic context the
statement that the bulk solution near the horizon does not depend upon
the details of the matter at large values of radial coordinate (boundary).
Indeed, within the attractor mechanism, the black hole horizon (IR region)
does not have any memory of the `initial conditions' (UV values of the
moduli) at the boundary. This is due to the existence of an infinite throat
in the extremal near horizon geometry. In the presence of one gauge field
(electric or magnetic) coupled with
one scalar, the extremal limit is a naked (null) singularity. The reason
is that the effective potential \cite{Goldstein:2005hq} does not have a
minimum at the horizon. However, when the effective potential has a minimum
at the horizon, there exist extremal solutions with finite
horizon area. In our case we are not able to define an effective potential,
but we have used, instead, the entropy function formalism. Therefore, the
attractor mechanism can be regarded as a `litmus test', which all extremal
black holes with finite area should pass (the horizon values of the moduli
should be finite).

In the presence of Chern-Simons term, there is no known concrete
expression for the effective potential (there is a non-trivial fibration
for $AdS_2$). Instead, we have used the entropy function and shown that
the horizon value of the scalar is fixed. What came as a surprise was the fact that
the modulus is fixed by the Chern-Simons coupling only. A simple
argument\footnote{We would like to thank Rob Myers for a discussion on this point.}
of why is so is that, by rescaling the metric, we can connect the (near horizon
extremal) solution in the absence of the modulus with the solution when
the modulus is turned on. The starting point is a finite area solution and so
we use the solution for $\zeta=\pm 1$ in the absence of the modulus. We then
obtain a finite area extremal near horizon solution for which the modulus is
fixed by the new Chern-Simons coupling. Therefore, in the presence of the moduli,
there exist finite area horizon solutions for any value of the Chern-Simons
coupling. We can also safely argue that the large ground state degeneracy
is stable against scalar field perturbations. However, as in the case when
the modulus is turned on, there also exists a critical value of the magnetic
field and so we expect similar physical interpretations as the ones
in \cite{D'Hoker:2010rz}.

Let us now comment on the second part of our paper. We have
computed the shear viscosity to entropy density ratio in the presence of the
most general bosonic four-derivative action (with one electromagnetic
field) with moduli dependent couplings.

In general, there are two types of first-order corrections
due to higher derivative terms. The entropy/area law is modified due to
the additional terms in the action and/or the modification of the area
due to the change of the metric at the horizon (the extra terms in
the action change the equations of motion). Therefore, one has to use
Wald formula to compute the entropy.

To compute the shear viscosity, we have used the method of \cite{ns1}. Since
we needed  just the near horizon data, the computations are not very complicated
even in the presence of higher derivative terms. We would like to
emphasize that there is no ambiguity in defining the overall coefficient
of the effective action and the results are consistent. Another important
observation is that, in the presence of a magnetic field, there is a
`preferential' direction in the plasma. A priori it is not clear that
the dual gravitational mode, $h_{xy}$, decouples. Indeed, that is not
the case when the momentum is non-zero --- however, in the hydrodynamic
limit we have the required decoupling \cite{Buchbinder:2008nf} and
this is what we needed for the computation of the corresponding
correlation function.

The shear viscosity to entropy density ratio is controlled by the
horizon values of the scalars and so, in the non-extremal case, an
operator deformation in the QFT will produce an interpolating non-trivial
flow in which the moduli approach the (IR) black hole
horizon. In the extremal case, though, the horizon moduli
values are fixed and so the shear viscosity to entropy density
ratio does not depend on the asymptotic values of the scalars. Therefore,
QFTs with different UV fixed points can flow to
the same IR fixed point.

In the presence of the moduli, $\eta/s$ also depends
only on the $c_1$ ($\sim (c-a)/c$, where $c$ and $a$ are the
central charges in the CFT) and $c_2$ (the coupling of
the stress tensor to the $U(1)$ current) coefficients. The other
two coefficients, $c_3$ and $c_4$, parametrize couplings of the
four-point function of the $U(1)$ currents and so they should
appear in the expressions of the charge density and conductivity
\cite{mps}.

It can be easily checked, that, when the scalars are turned off we recover
the results of \cite{mps}. Since \cite{mps} contains a very detailed
discussion on the relation with the previous results in the literature, we
do not want to repeat it here. We would like to point out just
that, in the extremal limit and in the presence of the magnetic field, we
also obtain a {\it deconfined} `plasma'  in the dual CFT. However, in
the presence of the magnetic field, $\eta/s$ depends on both coefficients,
$c_1$ and $c_2$ --- when the magnetic field vanishes, $\eta/s$ only
depends on $c_2$ and we recover the result of \cite{mps}. In $N=2$
supergravity, the bulk supersymmetry constraints all four derivative
couplings to be proportional to a single overall constant.

When the modulus is turned on, the extremal limit is more intriguing. The
dependence of $\eta/s$ on $c_2$ drops out and so $\eta/s$ depends only
on the central charges in the dual CFT. This is somehow similar
with the $N=2$ supergravity case mentioned above for which all the
coefficients are fixed by the central charges.

We close with some future directions.

Due to the presence of a Chern-Simons term, there is a more
interesting kind of instability\footnote{We would like to thank Hirosi
Ooguri for pointing out this to us.}, which can appear \cite{Nakamura:2009tf}.
This instability is caused by a non-normalizable mode of the AdS$_2$
factor in the near horizon geometry.\footnote{It was shown in
\cite{Nakamura:2009tf} that the near horizon analysis gives a
sufficient but not necessary condition. The reason is that, for the
Reissner-Nordstr\"om case studied in \cite{Nakamura:2009tf}, there
are unstable modes in the full bulk geometry that do not reduce to
normalizable modes in the near horizon limit.} This instability
appears for large enough Chern-Simons couplings and a {\it finite}
momentum.

It will be interesting to check \cite{DNS} if this kind of instability
appears also in our case. We would like to point out that the near horizon
geometry, when there are background electric and magnetic fields, is in
fact a fibered $AdS_2\times R^3$, not a direct product as in
\cite{Nakamura:2009tf}. In this case an analysis of the near horizon
geometry as in our paper (see, also, \cite{Astefanesei:2009sh}) is more appropriate.

Using arguments as in \cite{Banerjee:2010zd}, a computation of other
transport coefficients is also possible \cite{DNS1}.
However, since the near horizon data are not enough, a more involved analysis
is needed. Another reason of why the analysis is more complicated is that
the hydrodynamic limit does not commute with the limit of small magnetic
fields. Therefore, one expects drastic changes of the transport properties
of magnetized fluids. In this context, it will be interesting to study
how the attenuation and the speed of sound waves are affected by the
background magnetic field --- a similar analysis for a $(2+1)$-dimensional
plasma was presented in \cite{Buchbinder:2008nf}. It will also be
interesting to see if new dissipative transport coefficients can
appear in the effective hydrodynamic description of plasmas in external
magnetic fields \cite{Buchbinder:2009mk}.

\vspace{1.5cm}

\section*{Acknowledgments}
We would like to thank Stefan Theisen for fruitful discussions and
collaboration in the initial stage of this project. We would also
like to thank Alex Buchel, Dileep Jatkar, Per Kraus, Prem Kumar, Rob Myers, Hirosi Ooguri,
Bernard de'Wit for interesting discussions, and especially Ashoke Sen
for valuable discussions and comments on an earlier draft of this paper.
DA would like to thank ESI, Vienna for the hospitality during the
last stages of this work and the organizers of ESI Programme on
AdS Holography and the Quark-Gluon Plasma for a stimulating environment.
NB and SD acknowledge the hospitality of AEI, Potsdam at various
stages of this work. NB is also thankful to Swansea University for
the hospitality at the final stage of this work. The work of
NB is a part of research programme of FOM, which is financially
supported by the Netherlands Organization for Scientific Research (NWO).

\vspace{1cm}

\appendix

\noindent
{\Large {\bf Appendix}}

\section{Kubo formula and shear viscosity}

In this appendix we present a brief review of Kubo formula
in the context of relativistic hydrodynamics. Since we are
interested in computing the transport coefficients by using
the AdS/CFT duality, we follow closely the nice reviews
\cite{Son:2009zz} (see, also,
\cite{Ollitrault:2007du} for a good introduction in
relativistic hydrodynamics).

Current understanding is that the matter produced in the heavy-ion
experiments at RHIC behaves collectively like a fluid. That is,
the system rapidly thermalizes and comes into {\it local} thermal
equilibrium; then it evolves according to hydrodynamics until it
hadronizes and the temperature becomes smaller than the deconfinement
temperature.

Therefore, the relativistic hydrodynamics --- in short,
the Navier-Stokes equations and their relativistic
generalizations --- is the best method currently available
for modeling the spacetime evolution of sQGP.

Hydrodynamics can be understood as an effective theory, which
describes the dynamics at large length and time-scales. It
relies only on the assumption of local equilibrium.\footnote{One
does not have to make other assumptions
on the classical/quantum nature of the phenomena involved or the
type of particles/fields and their interactions.}

In order to describe essential non-equilibrium phenomena in heavy-ion
collision, a transport theory approach seems necessary. The first
important step to investigate small perturbations of a high
temperature hadronic matter is the linear response formalism.

If $l_{\mathrm{mfp}}$ is the mean free path --- the average distance
traversed between collisions by particles (of the
liquid) --- then, a perturbation of the liquid (in equilibrium)
will disappear at distances of order $l_{\mathrm{mfp}}$. However, the
fluctuations associated to some conservation laws can propagate
without being damped off. For example, a perturbation of
(mean) energy can propagate to infinity if there is no internal
friction ({\it viscosity}) of the fluid.

In the rest of this section we would like to explain why a measure
of $\eta/s$ is important and we also briefly review the Kubo's
formula, which is useful for computing the viscosity.

Unlike in standard thermodynamics where the system is in global
thermodynamic equilibrium (the intensive parameters, e.g. the pressure
($P$) and temperature ($T$), are constant throughout the volume)
and at rest, we are interested in systems whose pressure and
temperature vary with space and time, and which are not at rest.

Though, we request that the system is in local thermodynamic
equilibrium. That is,  the fluid  is characterized only by
local temperature and velocity fields that vary slowly on the
scale set by the temperature. Consequently, one can assume
the thermodynamic equilibrium in some neighborhood about
any point.

In hydrodynamics, it is useful to work with densities per unit
volume: the energy density $\epsilon\equiv U/V$ and the entropy
density $s\equiv S/V$. The usual relation for the energy can be
rewritten as $Ts=\epsilon+P$ --- note that all these densities are
intensive quantities.\footnote{All thermodynamic quantities
associated with a fluid element, e.g. $s, \epsilon,$ and $P$,
are defined in the rest frame and so they are Lorentz scalars
by construction.}

In hydrodynamics, we work directly with the equations of motion
and assume an expansion in derivatives so that
$\partial_\mu\ll l_{\mathrm{mfp}}^{-1}$. We express the stress tensor
through the temperature $T(x)$ and velocity $u^{\mu}(x)$.

For an ideal fluid, the hydrodynamic equations are equivalent
with the laws of conservation of energy and momentum:
\be
\partial_{\mu}T^{\mu\nu}=0\, , \,\,\,\,\,\,\,\,
T^{\mu\nu}=(\epsilon+P)u^{\mu}u^{\nu}+Pg^{\mu\nu}.
\ee
It is clear from the above expression that the momentum density
is $(\epsilon+P)\vec{v}$ and so, unlike the non-relativistic
fluid, the pressure contributes to the inertia of a relativistic
fluid.

At the next order we obtain the energy momentum tensor of
a relativistic viscous fluid whose conservation equation
reproduces the relativistic version of the Navier-Stokes
equation. The stress tensor at this order becomes
\be
T^{\mu\nu}= P g^{\mu\nu}+(\epsilon+P)u^{\mu}u^{\nu} - \sigma^{\mu\nu}\nn
\ee
with the dissipative part
\be
\sigma^{\mu\nu}=P^{\mu\alpha}P^{\nu\beta} \left[\eta \left(\partial_\alpha u_\beta+
\partial_\beta u_\alpha - \frac{2}{3}g_{\alpha\beta}\partial_\mu u^\mu\right) +
\varrho g_{\alpha\beta}\partial_\mu u^\mu\right].
\ee
Here, $P^{\mu\nu}=g^{\mu\nu}+u^{\mu}u^{\nu}$ is the usual projector
operator: for a fluid at rest it becomes
$P^{\mu\nu}=\mathrm{diag}(0,+1,+1,+1)$ and so it projects on space.

The numerical coefficient of the traceless part, $\eta$, is called the
shear viscosity. The numerical coefficient of the trace part, $\varrho$,
is called bulk viscosity.

To understand better the physical interpretation of the transport
coefficients and why a measure of $\eta/s$ (and $\varrho/s$) is
relevant (see, also, \cite{Mia:2009wj}), let us work in the fluid
rest frame $u^{\mu}=(1,0,0,0)$ and consider a small fluctuation
around the thermal equilibrium: $u^i\ll 1$, $u^0\backsimeq 1+O(u^iu^i)$,
$T=T+\delta T$, {\it etcetera}.

For the ideal fluid, we find the following perturbed stress tensor
at first order:
\be
\label{T}
T^{00}=\epsilon+\delta\epsilon\, , \,\,\,\,\,\, T^{0i}=(\epsilon+P)u^i\, ,
\,\,\,\,\,\, T^{ij}=(P+\delta P)\delta^{ij}.
\ee

The deviation from the ideal fluid stress tensor (at first order)
in the fluid rest frame (we use the properties of the projector) is
\ben
\delta T_{ij}&=&\sigma_{ij}=\eta \left(\partial_j u_i +
\partial_i u_j - \frac{2}{3}\delta_{ij}\partial_k u^k\right) +
\varrho \delta_{ij}\partial_k u^k \nn \\
&=&
\frac{\eta}{Ts} \left(\partial_j T_i^0+
\partial_i T_j^0 - \frac{2}{3}\delta_{ij}\partial_k T^{k0}\right) +
\frac{\varrho}{Ts} \delta_{ij}\partial_k T^{k0}.
\een
In the last expression we used (\ref{T}) and $Ts=\epsilon+P$.

Now it is clear that, since the temperature is the only relevant
energy scale (experimental parameter), the viscous terms are
characterized by the coefficients $\eta/s$ and $\varrho/s$ (medium
parameters). The physical interpretation is as follows:
the bulk viscosity encodes the resistance of the system to uniform
expansion and the shear viscosity controls the rate of the momentum
diffusion in the transverse direction to the flow.

For {\it conformal} fluids the energy-momentum tensor should be
traceless in flat space, which implies $\varrho=0$ and $\epsilon=3P$.

We have defined the dissipative coefficients as phenomenological
constants. However, it is important to find a way to compute them
directly from the microscopic theory. The computation of the shear
viscosity in quantum field theory relies on the Kubo formula --- the
dual models of CFTs are well suited to evaluate Kubo expressions.

Let us now briefly present Kubo formula and its physical
interpretation.

We are interested to understand the {\it response} of a
macroscopic quantum (many-body) system to a localized
disturbance, which is created by an applied external force.
Let us consider a local observable $O(\vec{x},t)$ (e.g., the
charge current or $T^{\mu\nu}$, which is a set of conserved
currents) and an external source that couples linearly to
the observable so that the new action is
\ben
S=S_0+\int dx O(x)J(x).
\een

The response, which is the change (from zero at the equilibrium)
in the expectation value of the observable induced by the source,
is linear in the source:
\ben
<O(x)>=\int dy \chi(x-y)J(y)\equiv \chi \centerdot J.
\een
The coefficient of proportionality is called `susceptibility' and
is nothing else than the retarded Green function of the physical
observable:
\ben
\chi(x-y)\equiv -\frac{i}{\hbar}\theta(x^0-y^0)<[O(x),O(y)]>.
\een
One can also Fourier transform to obtain that, indeed,
\ben
\chi(\omega,\vec{p})=\frac{<O(\omega,\vec{p})>}{J(\omega,\vec{p})}.
\een

Let us know find the Kubo formula for viscosity (see, e.g., \cite{Son:2009zz}).
We consider a perturbation of the metric so that the only non-zero component
$h_{12}(t)$ does not depend of space coordinates (it corresponds
to $\vec{p}=0$ in momentum space). Obviously, this perturbation
can not excite the temperature (which is a scalar) and the
velocity (which is a vector). One can easily compute the $12$-component
of the stress tensor (by using the covariant derivative) to obtain
\ben
\eta=-\lim_{\omega\to 0}\frac{1}{\omega}\mathrm{Im}\left(\int d^4x\, \theta(t)
e^{i\omega t}<[T_{12}(x),T_{12}(0) ]>  \right).
\een

\section{Coefficients in expansion (\ref{nonextremal}) at ${\cal O}(r-r_h)^2$}
For completeness, in this appendix we present the other coefficients that
appear in the near horizon expansion
of different fields in eq. (\ref{nonextremal}). They are
\ben
\label{BB}
u_2 &=& e^{\alpha  \varphi _h} \left(\frac{5 B^2}{3 r_h^4}+\frac{7
   q^2}{3}\right)+\frac{3}{4} z_1^2 r_h^2-2 \nn
\een
\ben
v_2&=&{1\over 288 \pi ^2 T^2}\bigg(e^{-2 \alpha  \phi _h}
\bigg(24 B^4 \zeta ^2 e^{\alpha  \phi _h}+24 \bigg(5 B^2+2 q^2\bigg)
e^{3 \alpha  \phi _h}-72 B^2 \zeta ^2\nn \\
&&+\bigg(6
   B^4 \bigg(\alpha ^2-4\bigg)-B^2 q^2 \bigg(9 \alpha ^2+8\bigg)+q^4
\bigg(3 \alpha ^2-4\bigg)\bigg) e^{4 \alpha  \phi _h}-144 e^{2 \alpha
   \phi _h}\bigg)\bigg)\nn
\een
\ben
w_2 &=& {e^{-4 \alpha  \phi _h}\over 288 \pi ^2 T^2}
\bigg(-36 B^4 \zeta ^4-24 B^2 \zeta ^2 \bigg(B^2+q^2\bigg)
e^{3 \alpha  \phi _h}+48 \bigg(q^2-2 B^2\bigg) e^{5
   \alpha  \phi _h}+288 B^2 \zeta ^2 e^{2 \alpha  \phi _h} \nn \\
&& +\bigg(-3 B^4 \bigg(\alpha ^2-4\bigg)-8 B^2 q^2+q^4 \bigg(3 \alpha ^2-4\bigg)\bigg)
   e^{6 \alpha  \phi _h}-144 e^{4 \alpha  \phi _h}\bigg)\nn
\een
\ben
z_2&=&-\frac{B \zeta  \left(B^2 \left(9 \zeta ^2
e^{-3 \alpha  \phi _h}+1\right)+5 \left(q^2-6
e^{-\alpha  \phi _h}\right)\right)}{6 \pi  T}
\nn
\een
\ben
p_1&=&\frac{B q \zeta  e^{-3 \alpha  \phi _h}
\left(\left(3 \alpha ^2+4\right) \left(B^2-q^2\right)
e^{3 \alpha  \phi _h}-12 B^2 \zeta ^2+24 e^{2
   \alpha  \phi _h}\right)}{96 \pi ^2 T^2}.
\nn
\een

\section{Details on entropy function when the modulus is turned on}

In the presence of the modulus, the attractor equation are:

\ben
\label{eoms}
-\frac{\beta \pi \sqrt{\upsilon_2}}{L \upsilon_1^2}
\left[4 B^2 L^2 e^{\alpha \varphi_h}+ \upsilon_1^2
(-4 L + 24 L^2 + 4 q^2 e^{\alpha \varphi_h} +
\upsilon_2 z1^2)\right]&=&0 \nn \\
\frac{\beta \pi}{L \upsilon_1 \sqrt{\upsilon_2}}
\left[4 B^2 L^2 e^{\alpha \varphi_h} - \upsilon_1^2
(-4 L + 24 L^2 + 4 q^2 e^{\alpha \varphi_h} + 3
\upsilon_2 z_1^2)\right]&=&0 \nn \\
\frac{\beta \pi \sqrt{\upsilon_2}}{L \upsilon_1^2}
\left[4 B^2 L^2 e^{\alpha \varphi_h} + \upsilon_1^2
 (-24 L^2 + 4 q^2 e^{\alpha \varphi_h} + \upsilon_2 z_1^2)\right] &=&0 \nn \\
\frac{2 \beta \pi}{L}\left[L \left(\Theta-4 B p_1^2 \zeta \right)
+\upsilon_1 \sqrt{\upsilon_2} (4 p_1 q e^{\alpha \varphi_h}
-\upsilon_2 z1)\right]&=&0 \nn \\
\frac{8 \beta \pi q}{L}\left[2 B L \zeta +\upsilon_1
\sqrt{\upsilon_2} z_1 e^{\alpha \varphi_h} \right] &=&0 \nn \\
2 \pi  \beta \left(8 B p_1 \zeta -\frac{4 \upsilon_1
\sqrt{\upsilon_2} q e^{\alpha \varphi_h}}{L}+Q\right)&=& 0 \nn \\
-2 \beta e^{\alpha \varphi_h} L \upsilon_1 \sqrt{\upsilon_2} \pi \bigg(-{2 B^2 \over
    \upsilon_1^2} + {2 q^2 \over L^2}\bigg) \alpha &=& 0 .\nn \\
\een

From the last equation, we observe that, in the presence of the modulus,
a non-trivial extremal black hole can be obtained if neither $q$ nor $B$
vanish. Following the same steps as in Section 3.2.2, we can obtain
the  near horizon geometry. First we compute the following quantities:

\ben
q &=& {e^{-\alpha \varphi_h} L (Q + 8 B p_1 \zeta) \over
 4 \upsilon_1 \sqrt{\upsilon_2}}\nn \\
z_1 &=& {L (p_1 Q + \Theta + 4 B p_1^2 \zeta) \over \upsilon_1 \upsilon_2^{{3 \over 2}}} \nn \\
p_1 &=& {-Q + \sqrt{
  Q^2 \pm 16 B (2 B E^{{\alpha \varphi_h \over2}} \upsilon_2 \mp \Theta) \zeta} \over
 8 B \zeta}\nn \\
 \upsilon_2 &=& {Q^2 - 16 B \Theta \zeta \over 32 e^{\alpha \varphi_h}(3 \upsilon_1^2 -
   B^2 e^{-{ \alpha \varphi_h\over 2}} (e^{{3 \alpha \varphi_h\over 2}} \pm \zeta)) }.
\een

With these results, the constraint (5th) relation becomes
\be
{B \over 36 \upsilon_1} e^{-\alpha \varphi_h/2}
(e^{{3 \alpha \varphi_h \over 2}} \pm \zeta)
 \sqrt{{3 \upsilon_1^2- B^2 e^{\alpha \varphi_h} \over 2}}=0
\ee
and we obtain the following expression for the entropy:
\be
S=\beta \upsilon_1 \pi \sqrt{{Q^2 - 16 B \Theta \zeta \over 2 e^{{\alpha \varphi_h \over 2}}
(3 \upsilon_1^2  e^{{\alpha \varphi_h \over 2}} - B^2 (e^{{3 \alpha \varphi_h \over 2}} \pm \zeta)) }}.
\ee

Now, let us rewrite the near horizon geometry in terms of the charges

\ben
q&=&{1 \over 6 \upsilon_1} e^{-\alpha \varphi_h/2}
\sqrt{{3 \upsilon_1^2- B^2 e^{\alpha \varphi_h} \over 2}}\nn \\
z_1&=&\pm{2 B e^{{\alpha \varphi_h} \over 2} \over 3 \upsilon_1
\sqrt{{Q^2 - 16 B \Theta \zeta \over 2 e^{{\alpha \varphi_h}
\over 2}(3 \upsilon_1^2 e^{{\alpha \varphi_h} \over 2}
- B^2 (e^{{3 \alpha \varphi_h \over 2}} \pm \zeta))}}}.
\een

By using the last equation of \ref{eoms} we obtain
\be
\upsilon_1=\pm {B e^{\alpha \varphi_h/2} \over \sqrt{2}}
\ee
and from the fifth relation of \ref{eoms}, we see that the only possibility
is a fixed horizon value of the scalar:
\be
\varphi_h= {2 \over 3} \ln{|\zeta|}.
\ee
What comes as a surprise is the fact that the horizon value of
the scalar is fixed by the Chern-Simons coupling. Consequently,
the entropy is
\be
S= \beta \pi \sqrt{{Q^2 -16 B \Theta \zeta \over 6 |\zeta|^{2/3}}}.
\ee

\section{Decoupling of $h_{xy}$ mode}
\label{hxy}
In what follows, we provide a detailed derivation of the decoupling 
of the dual gravitational mode $(4.1)$. We have explicitly checked that 
the $h_{xy} = e^{i t \omega +2 V(r)} \epsilon  \Phi (r)$ mode does not 
couple with any other field when the momentum vanishes. 

For two derivative gravity theory, this can be easily seen from the 
equations of motion (\ref{einstein}, \ref{scalar}, \ref{gauge}). However, 
we are interested in the most general four-derivative action (\ref{hdaction}). 
In this case, instead of writing the equations of motion in the 
presence of higher derivative terms, we will explicitly compute
the action up to order $\epsilon^2$. In this way, it can be explicitly 
checked that there is no coupling between $h_{xy}$ and the other fields.

Let us turn on the following perturbations of the metric 
\ben
g_{\alpha\beta}&=&g^{(0)}_{\alpha\beta}+ \epsilon h_{\alpha\beta} \nn \\
&=&\left(
\begin{array}{ccccc}
 \frac{1}{U(r)}+e^{i t \omega } \epsilon  \xi_1 (r) & e^{i t \omega } 
\epsilon  \xi_2 (r) & e^{i t \omega } \epsilon  \xi_3 (r) & e^{i t 
\omega } \epsilon  \xi_4 (r) & e^{i t \omega } \epsilon  \xi_5 (r) \\
 e^{i t \omega } \epsilon  \xi_2 (r) & e^{2 W(r)} Z(r)^2-U(r) & e^{i t \omega } \epsilon  \Upsilon (r) & 0 & e^{2 W(r)} Z(r)+e^{i t \omega } \epsilon  \upsilon (r) \\
 e^{i t \omega } \epsilon  \xi_3 (r) & e^{i t \omega } \epsilon  \Upsilon (r) & e^{2 V(r)} & e^{i t \omega +2 V(r)} \epsilon  \Phi (r) & e^{i t \omega } \epsilon  \chi (r) \\
e^{i t \omega } \epsilon  \xi_4 (r)  & 0 & e^{i t \omega +2 V(r)} \epsilon  \Phi (r) & e^{2 V(r)} & 0 \\
 e^{i t \omega } \epsilon  \xi_5 (r) & e^{2 W(r)} Z(r)+e^{i t \omega } \epsilon  \upsilon (r) & e^{i t \omega } \epsilon  \chi (r) & 0 & e^{2 W(r)}
\end{array}
\right)\nn \\
\een
and gauge gauge fields
\ben
A_{\alpha}&=&A^{(0)}_{\alpha} + \epsilon f_{\alpha} \nn \\
&=& \left(e^{i t \omega } \epsilon  a_r(r) , e^{i t \omega } \epsilon  a_t(r)-E(r) , \frac{B y}{2}+e^{i t \omega } \epsilon  a_x(r), -\frac{B x}{2}+e^{i t \omega } \epsilon  a_y(r),e^{i t \omega } \epsilon  a_z(r)+P(r)\right)\nn \\
\een
Here $g^{(0)}_{\alpha\beta}$ and $A^{(0)}_{\alpha}$ are the background 
metric (\ref{anz2}) and the background gauge field; $\epsilon$ dependent terms 
are the perturbations.

Using these field excitations, we can now compute the action.\footnote{ 
We use the Mathematica notebook for this computation. We emphasize that our 
system is symmetric in $x-$ and $y-$directions.}  The result is complicated, 
but for our purpose it is enough to pick up the $\Phi(r)$-dependent 
terms --- for concretness, let us write the $\Phi(r)$-dependent part 
of the action:

\ben
S_{\Phi, \Phi}&=&
\Phi(r)^2     \bigg[ {1 \over 4 U(r)} ((e^{-2 V(r)-W(r)} (-U(r) e^{2 W(r)} (4 B^2 e^{\alpha  \varphi (r)}+4 Q(r)^2 e^{4 V(r)+\alpha  \varphi (r)} \nn \\
&&-2 e^{4 V(r)} U''(r)  
-4
   e^{4 V(r)} U'(r) \left(2 V'(r)+W'(r)\right)+e^{4 V(r)+2 W(r)} Z'(r)^2+24 e^{4 V(r)}) \nn\\
&&+2 U(r)^2 e^{4 V(r)} (2 P'(r)^2 e^{\alpha
   \varphi (r)}+e^{2 W(r)} (4 V''(r)+4 V'(r) W'(r)+6 V'(r)^2  \nn\\
&&+2 W''(r)+2 W'(r)^2-\varphi '(r)^2))+14 \omega ^2 e^{4 V(r)+2
   W(r)})))\bigg]  \nn\\
&& +
\Phi(r) \Phi'(r) \bigg[  (2 e^{2 V(r)+W(r)} \left(U'(r)+U(r) \left(3 V'(r)+W'(r)\right)\right)  \nn\\
&&-(2 \alpha' e^{W(r)-2 V(r)} (-2 U(r) V'(r) (U(r) \left(B^2 c_2+c_1
   e^{4 V(r)+2 W(r)} Z'(r)^2\right)-2 c_1 U(r)^2 e^{4 V(r)}  \nn\\
&& \left(2 V''(r)+3 V'(r)^2+W'(r)^2\right)+2 c_1 \omega ^2 e^{4 V(r)})+2 c_1 U(r)
   e^{4 V(r)} U'(r)^2 V'(r) \nn\\
&& +c_1 e^{4 V(r)} U'(r) \left(2 U(r)^2 \left(V''(r)+3 V'(r)^2\right)+\omega ^2\right)))/U(r))\bigg] \nn
\een
\ben
&&+
\Phi(r)\Phi''(r)  \bigg[(2 U(r) e^{2 V(r)+W(r)}-4 c_1 \alpha' U(r) e^{2 V(r)+W(r)}(U'(r) V'(r)\nn \\
&& +2 U(r) \left(V''(r)+V'(r)^2\right)))\bigg] \nn\\
&&+\Phi'(r)^2  \bigg[( \alpha' e^{W(r)-2 V(r)} (U(r) (B^2 c_2  +2 c_1 e^{4 V(r)} U'(r) V'(r)-c_1 e^{4 V(r)+2 W(r)} Z'(r)^2)  \nn\\
&& +c_1 e^{4 V(r)} \left(U'(r)^2+4
   \omega ^2\right)+2 c_1 U(r)^2 e^{4 V(r)} \left(-2 V''(r)+V'(r)^2+W'(r)^2\right))\nn \\
&& +\frac{3}{2} U(r) e^{2 V(r)+W(r)})\bigg] \nn\\
&& + \Phi'(r)   \Phi''(r) \bigg[ ( 2 c_1 \alpha' U(r) e^{2 V(r)+W(r)} \left(U'(r)+4 U(r) V'(r)\right))\bigg]  \nn\\
&&+ \Phi''(r)^2  \bigg[2 c_1 \alpha' U(r)^2 e^{2 V(r)+W(r)}\bigg]
\een

Since there is no mixing between $\Phi$ and the other modes in the zero momentum 
limit, this mode remains massless and a computation of the related Green's 
function from the near-horizon data is still possible.

However, the other metric fluctuations namely $h_{xz}$ or $h_{yz}$ are not decoupled 
from the gauge field perturbations in the zero momentum limit (even in the absence of higher 
derivative terms) --- these coupled terms are proportionl to B or $\zeta$ (CS term). 
Therefore, these modes are not massless in the zero momentum limit and 
the near horizon geometry is not sufficient to compute the related two point 
correlation functions, i.e. $<[T_{xz},T_{xz}]>$ or $<[T_{yz},T_{yz}]>$ .




\end{document}